%% file: main.tex
\documentclass{article}

\PassOptionsToPackage{numbers, compress}{natbib}

\usepackage[final]{neurips_2022}

\usepackage[utf8]{inputenc} 
\usepackage[T1]{fontenc}    
\usepackage{hyperref}       
\usepackage{url}            
\usepackage{booktabs}       
\usepackage{amsfonts}       
\usepackage{nicefrac}       
\usepackage{microtype}      
\usepackage{xcolor}         
\usepackage{tcolorbox}
\usepackage{graphicx}
\usepackage{multirow}
\usepackage{makecell}
\usepackage{cancel}
\usepackage{bbm}
\usepackage{mathtools}
\usepackage{enumitem}
\usepackage{csquotes}
\usepackage{subcaption}
\usepackage{algorithm,eqparbox}
\usepackage{algpseudocode}
\usepackage{caption}
\usepackage{multirow}
\usepackage{makecell}
\usepackage{multibib}
\usepackage{chngpage}

\newcites{appendix}{References}

\usepackage{hyperref}
\hypersetup{
    colorlinks = true,
    linkcolor = red,
    anchorcolor = blue,
    citecolor = blue,
    filecolor = blue,
    urlcolor = blue
    }

\newcommand{\speaker}{\pi_{\theta}}
\newcommand{\listener}{\rho_{\phi}}
\newcommand{\probe}{\rho_{\omega^*}}
\newcommand{\listeneropt}{\rho^{*(\theta)}}

\newcommand{\Ladapt}{\mathcal{L}_\mathrm{adapt}}
\newcommand{\Ladaptestimate}{\hat{\mathcal{L}}_{\mathrm{adapt}}}

\newcommand{\Linfo}{\mathcal{L}_\mathrm{info}}
\newcommand{\Linfoestimate}{\hat{\mathcal{L}}_{\mathrm{info}}}

\renewcommand\paragraph{\textbf}

\title{Emergent Communication:
\\ Generalization and Overfitting in Lewis Games}

\author{
Mathieu Rita \\
INRIA, Paris\\
\texttt{mathieu.rita@inria.fr} \\
\And
Corentin Tallec \quad Paul Michel\thanks{This work was performed when Paul Michel was affiliated with Ecole Normale Supérieure PSL.} \quad Jean-Bastien Grill\\
DeepMind \\
\texttt{[corentint,paulmiche,jbgrill]@deepmind.com} \\
\And
Olivier Pietquin \\
Google Research, Brain Team\\
\texttt{pietquin@google.com}
\And
Emmanuel Dupoux \\
EHESS,ENS-PSL,CNRS,INRIA\\
Meta AI Research \\
\texttt{emmanuel.dupoux@gmail.com}
\And
Florian Strub \\
DeepMind \\
\texttt{fstrub@deepmind.com} \\
}

\begin{document}

\maketitle

\begin{abstract}
  Lewis signaling games are a class of simple communication games for simulating the emergence of language. In these games, two agents must agree on a communication protocol in order to solve a cooperative task. Previous work has shown that agents trained to play this game with reinforcement learning tend to develop languages that display undesirable properties from a linguistic point of view (lack of generalization, lack of compositionality, etc). In this paper, we aim to provide better understanding of this phenomenon by analytically studying the learning problem in Lewis games. As a core contribution, we demonstrate that the standard objective in Lewis games can be decomposed in two components: a co-adaptation loss and an information loss. This decomposition enables us to surface two potential sources of overfitting, which we show may undermine the emergence of a structured communication protocol. In particular, when we control for overfitting on the co-adaptation loss, we recover desired properties in the emergent languages: they are more compositional and  generalize better.
\end{abstract}

\section{Introduction}


Understanding the dynamics of language evolution has been a challenging if not controversial research topic in the language sciences ~\citep{Harnad1976OriginsAE,christiansen2003language}. Given that the very first human language cannot be unearthed from fossils~\citep{bickerton2007language}, computational models have been designed to simulate the emergence of a structured language within a controlled environment. In this line of work, Lewis signaling games~\citep{lewis1969convention} are among the most widespread playground environments to model language emergence: they are inherently simple, yet they exhibit a rich set of communication behaviors~\citep{crawford1982strategic,skyrms2010signals}. Therefore, understanding Lewis games dynamics may shed light on the prerequisites of language emergence.

In their original form, Lewis signaling games involve two agents: a speaker and a listener. The speaker observes a random state from its environment, e.g. an image, and sends a signal to the listener. The listener then undertakes an action based on this signal. Finally, both agents are equally rewarded based on the outcome of the listener's action. The resolution of this cooperative two-player game requires the emergence of a shared protocol between the agents~\citep{lewis1969convention,crawford1982strategic}. 
%
%
One way to model the emergence of such protocol is to give the agents the capacity to learn. The agents, and therefore, the communication protocol, are shaped by a sequence of trials and errors over multiple games~\citep{wagner2003progress,kirby2002emergence,steels1997synthetic,skyrms2010signals}. This learning-centric approach allows for a fine analysis of the language emergence dynamics~\citep{skyrms2010signals,huttegger2014some}. It also raises challenging learning-specific questions: What are the inductive biases present in the agent architecture and loss function that shape the emergent language ~\citep{kirby2001spontaneous}? How do agents generalize from their training set? Is the resulting language compositional~\citep{brighton2006}? What is the impact of overfitting~\citep{lazaridou2018emergence}?

Recently, there has been a resurgence of interest for such learning-based approaches following advances in machine learning~\citep{lazaridou2020emergent}. In these approaches, the speakers and listeners are modeled as deep reinforcement learning agents optimized to solve instances of the Lewis games~\citep{lazaridou2018emergence,havrylov2017emergence,ren2020compositional,li2019ease,harding-graesser-etal-2019-emergent}. The vast majority of these works explore Lewis games from an empirical perspective. However, some of the recent experimental results are at odds with experimental findings from the linguistics literature. For instance, the emergent protocols lack interpretability~\citep{kottur-etal-2017-natural}, generalization does not always correlate with language compositionality~\citep{chaabouni2020compositionality}, successful strategies are not naturally adopted in populations~\citep{rita2022on,chaabouni2022emergent}, and anti-efficient communication may even emerge~\citep{chaabouni2019anti}. It is unclear whether those empirical observations result from a learning failure, e.g. optimization problems, overfitting, or whether they are symptomatic of more fundamental limitations of Lewis games for modeling language emergence, e.g. lack of embodiment~\citep{harnad1990symbol,barsalou2008grounded,ossenkopf2022which,kalinowska2022situated}. Overall, it is crucial to establish new analytical insight to analyze Lewis games in the learning setting.

In this paper, we introduce such an analytical framework to diagnose the learning dynamics of deep reinforcement learning agents in Lewis signaling games. As a core contribution, we demonstrate under mild assumptions that the loss of the speaker and listener can be decomposed into two components when resolving Lewis signaling games: (i) an \emph{information loss} that maximizes the mutual information between the observed states and speaker messages; (ii) a \emph{co-adaptation loss} that aligns the speaker and listener's interpretation of the messages
(Section~\ref{loss_decompo}). Based on this decomposition, we empirically examine the evolution of these two losses during the learning process~(Section~\ref{sec:emp_results}). In particular, we identify an overfitting problem in the co-adaptation loss between the agents which undermines the emergence of structured language. We then show that the standard setup used in the deep language emergence literature consistently suffers from this overfitting issue (Section~\ref{sec:visualizing}). This realization explains some of the contradictory observations~\citep{chaabouni2020compositionality} and experimental choices from past works~\citep{ren2020compositional,li2019ease,rita2022on}. Finally, we explore regularization methods to tackle this co-adaptation overfitting. We observe that reducing the co-adaptation overfitting allows for developing a more structured communication protocol (Section~\ref{sec:expe_2}).

All in all, our contributions are three-fold: (i) we provide a formal description of Lewis games from a learning standpoint (Section~\ref{sec:decompo_enonce}); (ii) we apply this framework in experiments to show that degenerate results are primarily due to overfitting in the co-adaptation component of the game (Section~\ref{sec:visualizing}) ; (iii) we propose natural ways of tackling this overfitting issue and show that, when we control the receiver's level of convergence, we obtain a well-structured emergent protocol (Section
~\ref{sec:expe_2}).

\section{Analyzing Lewis Games}
\label{loss_decompo}

We show that Lewis games' objective decomposes into two terms: (i) an information loss that measures whether each message refers to a unique input; (ii) a co-adaptation loss that quantifies the alignment of the speaker's and listener's interpretation of the messages. 

For simplicity and to ease the reader's intuition, we focus on the reconstruction variant of Lewis games with agents optimizing the reconstruction log-likelihood in the main paper. In Appendix~\ref{app:sec:lewis_demo}, we show that our analysis extends to a broader of Lewis signaling games, e.g. discrimination games~\citep{chaabouni2022emergent, mu2021emergent, dessi2021interpretable, guo2021expressivity, ren2020compositional, lazaridou2018emergence, lazaridou2016multi, havrylov2017emergence, li2019ease, Lowe2020On}, and to a general form of reward that covers the rewards commonly used in emergent communication, e.g. log-likelihood~\citep{chaabouni2019anti, chaabouni2020compositionality, kharitonov2020entropy, rita2022on, rita-etal-2020-lazimpa, chaabouni2021communicating}, accuracy reward~\citep{ren2020compositional, lazaridou2018emergence, lazaridou2016multi, kottur-etal-2017-natural, li2019ease, harding-graesser-etal-2019-emergent, evtimova2018emergent}.

\subsection{Background: Lewis Reconstruction Games}
\label{game_descri}


\paragraph{Game formalism} In reconstruction Lewis games, a speaker observes a random object of its environment. The speaker then sends a descriptive message, which a second agent, the listener, uses to reconstruct the object. The success of the game is quantified by how well the original object is reconstructed~\citep{kharitonov2020entropy,chaabouni2019anti,rita-etal-2020-lazimpa,rita2022on}.
%
Formally, the speaker is parameterized by $\theta$ and the listener is parameterized by $\phi$. The observed object denoted by $x$ is selected from a set of objects denoted by $\mathcal{X}$. We denote by $X$ the random variable characterizing $x$, sampled from distribution $p$. The intermediate message sent by the speaker $m$ belongs to the set of all potential messages $\mathcal{M}$. The speaker follows a policy $\speaker$ which samples a message $m$ with probability $\speaker(m|x)$ conditioned on object $x$. We denote by $M_{\theta}$ the random variable characterizing the message $m$, sampled from $\speaker(\cdot | X)$. We denote by $\pi_\theta(m) = \sum_x \pi_\theta(m | x) p(x)$ the marginal probability of a message given policy $\pi_\theta$. Given a message $m$, the listener outputs a probability distribution over inputs $\listener(\cdot|m)$, and the probability of reconstructing the entire object $x$ given $m$ is thus $\listener(x|m)$.

\paragraph{Game objectives} In reconstruction games, the speaker and listener minimize the negative log likelihood of the reconstructed object. Both agents thus optimize the objective:
\begin{align}
    \mathcal{L}_{\theta,\phi} &= - \mathbb{E}_{x \sim p, m \sim \speaker(\cdot|x)}[\log\listener(x|m)],
\end{align}
where optimizing the speaker is a reinforcement learning problem whose parameters $\theta$ are optimized using policy gradient~\citep{sutton2000policy} and optimizing the listener is a supervised learning problem whose parameters $\phi$ are optimized with gradient descent. In our theoretical analysis, we consider that agents are not regularized. In practice, regularizations, e.g. entropy regularization~\citep{pmlr-v48-mniha16}, may be added to the game objective but it does not alter our main conclusions.


\subsection{Building Intuition on the Lewis Reconstruction Game Learning Dynamics}
\label{sec:intuition}

To get a better intuition of the dynamic of Lewis reconstruction games, we can analyze the form taken by the optimal listener, given speaker $\pi_\theta$.
In what follows, we use sub-script $\theta$ to denote an explicit dependency of the policy on parameters $\theta$, e.g. a policy parameterized with a neural network. Conversely, the use of super-script $*(\theta)$ corresponds to an implicit dependency of the policy on parameters $\theta$.
As shown in Appendix~\ref{app:sec:lewis_reco_log_reward}, given a message $m$, the optimal listener's distribution $\listeneropt(\cdot|m)$
can be written in closed-form:
\begin{align}
    \listeneropt(x|m) \colon\hspace{-.65em}=
    \frac{p(x)\speaker(m|x)}{\sum_{x' \in \mathcal{X}}p(x')\speaker(m|x')} \cdot
\label{inverse_bayes}
\end{align}
Here, $\listeneropt$ does not depend on $\phi$, but implicitly depends on $\theta$, as it is the optimal listener \emph{given a policy parameterized by $\theta$.} At each update, the listener $\listener$ gets closer to its optimum $\listeneropt(\cdot|m)$ . 
If we suppose that the listener perfectly fits $\listeneropt(\cdot|m)$ at any moment, the loss becomes:

\begin{align}
    \mathcal{L}_{\theta,\phi} = - \mathbb{E}_{x \sim p, m \sim \speaker(\cdot|x)}[\log \listeneropt(x|m)] = \mathcal{H}(X|M_{\theta}) =  -  I(X;M_{\theta}) + \mathcal{H}(X)
    \label{opt_listener_case}
\end{align}

where $\mathcal{H}(X|M_{\theta})$ is the conditional entropy of $X$ conditioned on $M_{\theta}$ and $I(X;M_{\theta})$ is the mutual information between $X$ and $M_{\theta}$.
Thus, if the listener is optimal  at every point in time, the speaker's task merely becomes the construction of a message protocol that maximizes the mutual information between objects and messages, i.e. the construction of an unambiguous message protocol. 


In practice, the listener never perfectly fits the optimum. In the following, we elucidate the effect of this gap between the listener and its optimum on the dynamics of the game.


\subsection{Analytical Result: The Lewis Games Loss Decomposition}
\label{sec:decompo_enonce}





\begin{tcolorbox}

In cooperative Lewis games, the agents' loss can be decomposed into two terms:

\begin{align}
    \mathcal{L}_{\theta,\phi} =  \Linfo + \Ladapt,
    \label{eq:SGTD}
\end{align}

\begin{itemize}[leftmargin=*]
    \item An \textbf{information term} $\Linfo$ quantifies the degree of ambiguity of the language protocol. It is minimal when each message refers to a unique object;
    \item A \textbf{co-adaptation term} $\Ladapt$ quantifies the gap between the listener and its optimum: the speaker's posterior distribution. This co-adaptive term is optimized both by the speaker and the listener. When the listener is optimal, this co-adaptation objective is zeroed.
\end{itemize}

In particular, the decomposition takes the following form in the Lewis reconstruction game:

\begin{align}
    \mathcal{L}_{\theta,\phi} =   \underbrace{\mathcal{H}(X|M_{\theta})}_{\Linfo} + \underbrace{\mathbb{E}_{m\sim \speaker}D_{KL}(\listeneropt(\cdot|m)||\listener(\cdot|m))}_{\Ladapt},
    \label{eq:SGTD}
\end{align}

\end{tcolorbox}

The proof of the decomposition is provided in Appendix~\ref{app:sec:lewis_demo}. Appendix~\ref{app:sec:lewis_demo} provides the proof for the reconstruction log-likelihood reward and extends to a broader class of Lewis signaling games, e.g. discrimination games, and general cooperative rewards covering usual emergent communication rewards, e.g. the accuracy reward. 
This decomposition gives us insights on the game dynamics and the constraints that shape languages in the game with neural agents:

\paragraph{The information loss} $\Linfo$ captures the speaker’s intrinsic objective: to develop an unambiguous protocol. $\Linfo$ is minimal, equals to $0$, when the communication protocol is unambiguous, i.e. every message from the speaker's policy $\speaker$ refers to a unique object. Conversely, $\Linfo$ is maximal, equal to $\mathcal{H}(X)$, when the message protocol is fully ambiguous, and $X$ and $M_{\theta}$ are independent variables.


\paragraph{The co-adaptation loss} $\Ladapt$ is specific to learning agents. This loss measures how far the listener $\listener$ is from its optimum $\listeneropt$. If $\Ladapt=0$, the listener and its optimum coincide. $\Ladapt$ has the particularity to be optimized by the two agents. From the listener's side, it merely corresponds to the optimization of its supervised task. From the speaker's side, it brings out that the speaker must adapt its language to the listener in addition to build an unambiguous message protocol. In other words, the co-adaptation loss pushes the speaker to develop a language that can be easily recognized by listeners. This pressure diminishes as the listener approaches its optimum. 

From a practical perspective, Equation~\eqref{eq:SGTD} yields the following individual gradients:
\begin{align}
\label{eq:grad_decomposition}
\left\{
    \begin{array}{ll}
        \nabla_{\theta}\mathcal{L}_{\theta} &= - \nabla_{\theta}I(X,M_{\theta}) + \nabla_{\theta} \mathbb{E}_{m\sim \speaker}D_{KL}(\listeneropt(\cdot|m)||\listener(\cdot|m)) \\
        \nabla_{\phi} \mathcal{L}_{\phi} &= \text{\hspace{2.53cm}} \nabla_{\phi} \mathbb{E}_{m\sim \speaker}D_{KL}(\listeneropt(\cdot|m)||\listener(\cdot|m)),
    \end{array}
\right.
\end{align}
where the listener only receives gradients from the co-adaptation term, and the speaker receives gradients from both terms.

This loss decomposition also finds echoes in the cognitive science literature in the form of an expressivity vs. learnability trade-off \citep{smith2003iterated}; see Section \ref{sec:related} for a detailed discussion.

\subsection{Generalization Gaps in Lewis Reconstruction Games}


We explore another facet of the loss decomposition that arises from learning. As agents are trained on partial views of their environment, it opens questions of overfitting and generalization to unseen objects.
As is customary in machine learning, we consider agents trained on a fixed, finite sample from the data distribution: the training set. Let us denote by $p_{\mathrm{train}}$ the \emph{empirical} object distribution over the training set and $X^{\mathrm{train}}$ an object sampled from $p_{\mathrm{train}}$. Similarly let $M_{\theta}^{train}$ denote a message sampled from $\speaker(.|X
^{\mathrm{train}})$, $\pi_\theta^\mathrm{train}(m) = \sum_x \pi_\theta(m|x) p_\mathrm{train}(x)$ the marginal probability of a message on the training set, and  $\listeneropt_{\mathrm{train}}(x|m)=\frac{p_{\mathrm{train}}(x)\speaker(m|x)}{\sum_{x \in \mathcal{X}}p_{\mathrm{train}}(x)\speaker(m|x)}$ the speaker's posterior distribution with respect to the prior distribution $p_{\mathrm{train}}$. The training loss can be written as follow:
\vspace{0.5em}
\begin{align*}
    \mathcal{L}_{\theta,\phi}^{\mathrm{train}} & = - \mathbb{E}_{x \sim p_{\mathrm{train}}, m \sim \speaker(\cdot|x)}[\log\listener(x|m)] \\
    & = \underbrace{\mathcal{H}(X^{\mathrm{train}}|{M^{\mathrm{train}}_{\theta}})}_{\Linfo^{\mathrm{train}}} +  \underbrace{\mathbb{E}_{m\sim \pi^\mathrm{train}_{\theta}}D_{KL}(\listeneropt_{\mathrm{train}}(\cdot|m)||\listener(\cdot|m))\cdot}_{\Ladapt^{\mathrm{train}}}
\end{align*}

Decomposing the gap between $\mathcal{L}^{\mathrm{train}}_{\theta,\phi}$ and $\mathcal{L}_{\theta,\phi}$ uncovers two sources of overfitting:
\begin{align}
    \mathcal{L}^{\mathrm{train}}_{\theta,\phi} & = \mathcal{L}_{\theta,\phi} 
    + \underbrace{\Linfo^{\mathrm{train}}-\Linfo}_{\text{information overfitting}}  + \underbrace{\Ladapt^{\mathrm{train}}-\Ladapt}_{\text{co-adaptation overfitting}}\cdot
\end{align}
Intuitively, \textit{information overfitting} occurs when the speaker only develops an unambiguous language on the training set, but ambiguities remain on the total dataset. \textit{Co-adaptation overfitting} occurs when the two agents agree on a common communication protocol on the training data, but not on all data.


\section{Method}

This section gathers the methodological tools required to empirically study the loss decomposition. 

\subsection{Probing the Information and Co-adaptation Losses}
\label{sec:probing}

\begin{figure}[ht]
\centering
\includegraphics[width=0.95\textwidth]{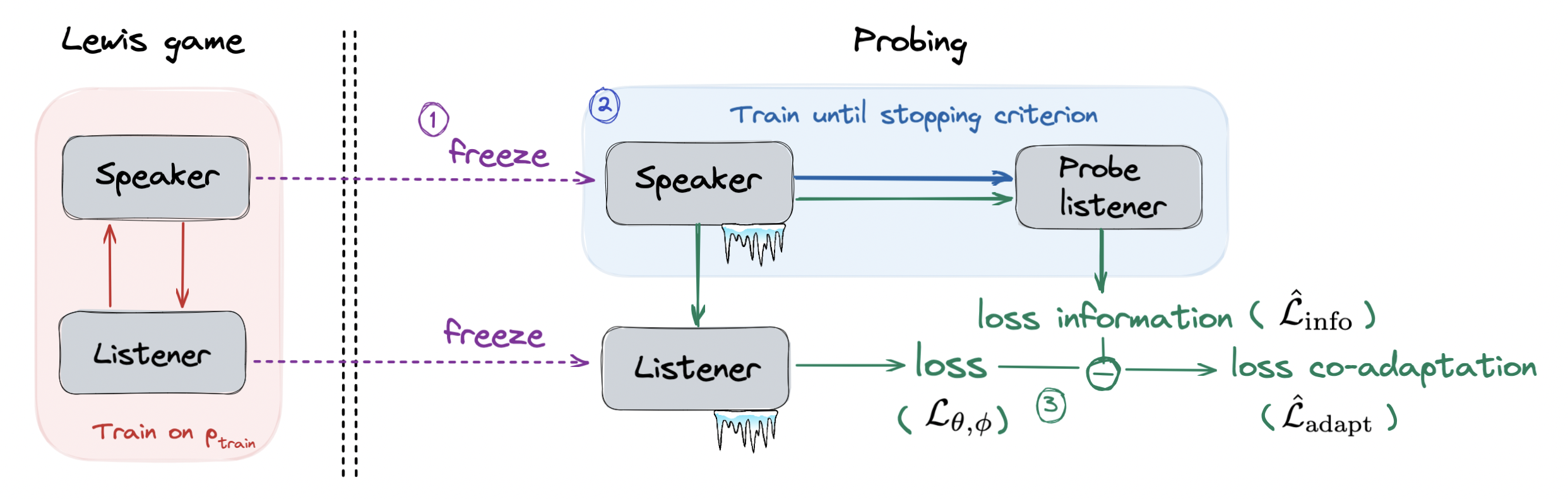}
\caption{Probing method: (1) the speaker and listener are frozen and the probe listener is initialized. (2) the probe listener is trained on $p_{\mathrm{train}}$ (resp. $p$) with the speaker's messages until convergence; (3) The speaker takes inputs from $p_{\mathrm{train}}$ (resp. $p_{\mathrm{test}}$) and messages the probe listener and the listener. The resulting loss of the probe listener is $\Linfoestimate$, and the loss of the listener is used to estimate $\Ladaptestimate$. 
}
\vskip -0.5em
\label{fig:prob_scheme}
\end{figure}

Computing $\Linfo$ and $\Ladapt$ directly necessitates estimating the posterior distribution of the speaker, $\listeneropt(.\mid m)$. Doing so requires summing over all $\mathcal X$ which is intractable. 
Fortunately, deep models are large enough so that they can perfectly solve their task on their train set. We can leverage this fact to compute empirical estimates $\Linfoestimate$ and $\Ladaptestimate$ of $\Linfo$ and $\Ladapt$ respectively by using an auxiliary listener trained to optimality. 

We here detail an empirical probing mechanism to obtain estimates $\Linfoestimate$ and $\Ladaptestimate$ given speaker $\speaker$ and listener $\listener$. 
%
%
As noted in Equation~\ref{inverse_bayes}, the posterior $\listeneropt$ also corresponds to the optimal listener. Therefore, we obtain an estimate of the posterior by training a listener to optimality, and use this optimal listener to decompose the loss. In practice, to obtain this optimal listener, we freeze speaker $\speaker$ and listener $\listener$ and initialize a new, auxiliary listener from scratch, which we refer to as the \emph{probe} listener. As illustrated in Figure~\ref{fig:prob_scheme}, the probe listener is trained to reconstruct object $x$ from message $m$, with $x$ drawn from distribution $p$ or $p_\mathrm{train}$ and $m$ sampled according to the frozen speaker policy $\speaker(.|x)$, until a stopping criterion is met. We then distinguish between the train and test estimates:
\begin{alignat}{1}
\begin{split}
\Linfoestimate^\mathrm{train} &= -\mathbb{E}_{x \sim p_{\mathrm{train}}, m 
\sim \speaker(\cdot|x)}[\log\probe^{\mathrm{train}}(x|m)] 
\\
\Ladaptestimate^\mathrm{train} &= -\mathbb{E}_{x \sim p_{\mathrm{train}}, m
\sim \speaker(\cdot|x)}[\log\listener(x|m)] - \Linfoestimate^\mathrm{train}
\end{split}
\end{alignat}
and,
\begin{alignat}{1}
\begin{split}
\Linfoestimate^\mathrm{test} &= -\mathbb{E}_{x \sim p_{\mathrm{test}}, m 
\sim \speaker(\cdot|x)}[\log{\probe}(x|m)] 
\\
\Ladaptestimate^\mathrm{test} &= -\mathbb{E}_{x \sim p_{\mathrm{test}}, m
\sim \speaker(\cdot|x)}[\log\listener(x|m)] - \Linfoestimate^\mathrm{test}
\end{split}
\end{alignat}

where $\probe^{\mathrm{train}}$ and $\probe$ are the probe listeners trained over distributions $p_{\mathrm{train}}$ and $p$ respectively.\footnote{The estimate 
is trained on $p$, the \emph{full} distribution of objects, and not $p_\mathrm{test}$. Training on $p_\mathrm{test}$ would result in an optimal listener overfitting on the test set, which would results in bad estimates of the mutual information.} 
Note that this probing mechanism, while tractable, is computationally costly as it necessitates training a new probe listener to convergence, and so we only use it as a valuable diagnosis tool.

\subsection{Balancing the Information and Co-adaptation Terms}
\label{sec:task_balance_method}

As explained in Section \ref{sec:decompo_enonce}, the information loss alone is sufficient for the speaker to develop an unambiguous language. This begets the question: does the co-adaptation loss have any bearing on the emergent language at all? We elucidate this question by balancing the weight of the co-adaptation term in the decomposition. By using the probing method described above, we build the following training loss:
\begin{align}
    \mathcal{L}_{\theta}(\alpha) &= (1-\alpha) \times \Linfoestimate^{\mathrm{train}} +  \alpha \times \Ladaptestimate^{\mathrm{train}} \qquad \text{where} \quad \alpha \in [0;0.5]
\end{align}

Hence, $\alpha$ balances the two speaker objectives (up to an approximation error). When $\alpha=0.5$, the loss falls back to the classic setting. When $\alpha=0$, the co-adaptation term is removed on the speaker side; note that the Lewis game can still be solved since the listener still optimizes the co-adaptation term. We experimentally analyse the effect of $\alpha$ on resulting languages in Section~\ref{sec:visualizing}.
In Appendix \ref{app:method_comp}, we describe how we build the balanced loss and explain why $\alpha$ should be bounded by 0.5.

\subsection{Controlling the Listener's Co-adaptation Loss Level of Convergence}
\label{sec:method_listener_cv_rate}

As mentioned in \ref{sec:intuition}, the influence of $\Ladapt$ on the co-adaptation term in the speaker's loss is modulated by the listener's level of convergence to its optimum. To understand the effect of this co-adaptation, we decouple the speaker and listener training and train the listener via three procedures:

\textbf{Continuous listener} The listener is continuously trained, jointly with the speaker. This is the standard setting in the emergent communication literature, and serves to report the baseline behavior.

\paragraph{Partial listener} The listener is re-initialized \emph{after each} of the speaker's update and trained on the training set for $N_{step}$ before updating the speaker again. This baseline enables fine-grained analysis of the influence of under-training (low $N_{step}$) and over-training (large $N_{step}$) the listener.

\paragraph{Early stopping listener} The listener is also re-initialized \emph{after each} of the speaker's update but is now trained until an early stopping criterion is met on the validation set. This allows us to get the best estimate of the posterior $\listeneropt(.|m)$ at each update. 
This can be seen as a variant of the partial listener with an adaptive number of steps $N_{step}$.


\section{Experimental settings}
\label{sec:exp_set}


\subsection{Game description}
\label{expe_set:reconstruction_game}
\vspace{-0.5em}

Unless specified, all our experiments are run on the reconstruction game defined in Section~\ref{game_descri}. 
Experiments are run over 6 seeds and reach $>99\%$ training reconstruction scores unless otherwise stated. Our implementation is based on the EGG toolkit~\citep{kharitonov2019egg} and the code is  available at \url{https://github.com/MathieuRita/Population}.

\paragraph{Environment} We consider objects $x =: (x_{1},...,x_{K}) \in \mathcal{X}=:\mathcal{X}_{1}\times ... \times \mathcal{X}_{K}$ characterized by $K$ attributes where attribute $i$ may take $|\mathcal{X}_{i}|$ different values. By design, this synthetic environment allows us to test the ability of agents to refer to unseen objects by communicating their attributes~\citep{baroni2020,kottur-etal-2017-natural}.
Each object is the concatenation of one-hot representations of the attributes $(x_{i})_{1 \leq i \leq K}$. Objects have $K=6$ attributes, each taking $10$ different values, for a total of $1$ million objects. Training, validation and test sets are randomly drawn from this pool of objects (uniformly and without overlap), and are respectively composed of $4000$, $1000$ and $1000$ elements. Thus, the agents only have access to a small fraction ($<1\%$) of the environment, making the generalization problem challenging.

\paragraph{Communication channel} Messages $m=:(m_{j})_{j=1}^{T} \in \mathcal{M} =: \mathcal{V}^{T}$ are sequences of $T$ tokens where each token is taken from a finite vocabulary $\mathcal{V}$, finishing by a hard-coded end-of-sentence token \texttt{EoS}. In our experiments, messages have maximum length $T=10$ and symbols are taken from a vocabulary of size $|\mathcal{V}|=10$ to prevent a bottleneck in the communication channel.


\paragraph{Speaker model} The speaker follows a recurrent policy: given an input object $x$, it samples for all $t \in [1,T]$ a token $m_{t}$ with probability $\speaker(m_{t}|m_{<t},x)$. The speaker takes in the object $x$ as a vector of size $K\times|\mathcal{X}_{.}|$ and passes it through a linear layer of size $128$ to obtain an object embedding, used to initialize a LSTM~\citep{hochreiter1997long} of size $128$ with layer normalization~\citep{ba2016layer}. At each time step, the LSTM's output is fed into a linear layer of size $|\mathcal{V}|$, followed by a softmax, to produce $\speaker(m_t|m_{<t},x)$

\paragraph{Listener model} Given a message $m=(m_{1},...,m_{T})$, the listener outputs for each attribute $k$ a probability distribution over the $|\mathcal{X}_{k}|$ values: $\listener^{k}(\cdot|m)$. The probability of reconstructing the entire object $x$ given $m$ is then $\listener(x|m):=\prod_k \listener^k(x^k|m)$. The listener passes each message $m_t$ through an embedding layer of dimension $128$ followed by a LSTM with layernorm of size $128$. The final recurrent state $h^\mathrm{l}_T$ is passed through $K$ linear projections of size $|\mathcal{X}_{.}|$, each followed by a softmax, providing $K$ independent probability distributions of sizes $|\mathcal{X}_{.}|$ to predict each attribute of $x$.

\paragraph{Optimization} The agents are optimized using \texttt{Adam}~\citep{adam-opt} with a learning rate of $5\cdot 10^{-4}$, $\beta_1=0.9$ and $\beta_2=0.999$ and a batch size of $1024$. For the speaker we use policy gradient~\citep{sutton2000policy}, with a baseline computed as the average reward within the minibatch, and an entropy regularization of $0.01$ to the speaker's loss~\citep{williams1991function}. In all experiments, we select the best models by early stopping.

\vspace{-0.5em}
\subsection{Evaluating emergent languages properties}
\label{expe_set:metrics}
\vspace{-0.5em}

\paragraph{Generalization} We measure generalization by computing the average test reconstruction score over all the attributes of a probe listener trained on the training set using an early stopping criterion on the validation set. Indeed, the trained listener $\listener$ may overfit to the training set, and so using it may under-estimate. Using a separate listener removes this bias.

\paragraph{Compositionality} Compositionality is a fundamental feature of natural language often seen as a precondition to generalize~\citep{bickerton2014more,sep-compositionality,townsend2018compositionality}. We assess the compositionality by computing the topographic similarity~\citep{brighton2006,lazaridou2018emergence}. It is defined as the Spearman correlation~\citep{kokoska2000crc,2020SciPy-NMeth} between the distance in input space, i.e. the average number of common attributes, and the distance in message space, i.e. the  edit-distance between the corresponding messages~\citep{levenshtein1966binary}. 
%
%
As we here deal with large object space and stochastic policies, we
use a bootstrapped estimate of topographic similarity as in~\citep{koehn2004statistical} to get reliable numbers. We
sub-sample $1000$ elements $x$ from the object space $\mathcal{X}$, and sample the corresponding message $m$ from the speaker's policy $\speaker(\cdot|x)$. We compute the topographic similarity for this batch of $1000$ pairs $(x,m)$. We repeat this protocol $100$ times and take the mean to measure compositionality.

\vspace{-0.5em}
\section{Empirical results}
\label{sec:emp_results}

\vspace{-0.3em}
\subsection{Visualizing the loss decomposition dynamics}
\label{sec:visualizing}
\vspace{-0.3em}


\begin{figure*}[th!]
\hfill
\begin{subfigure}[t]{0.33\textwidth}
\centering
\includegraphics[width=\textwidth]{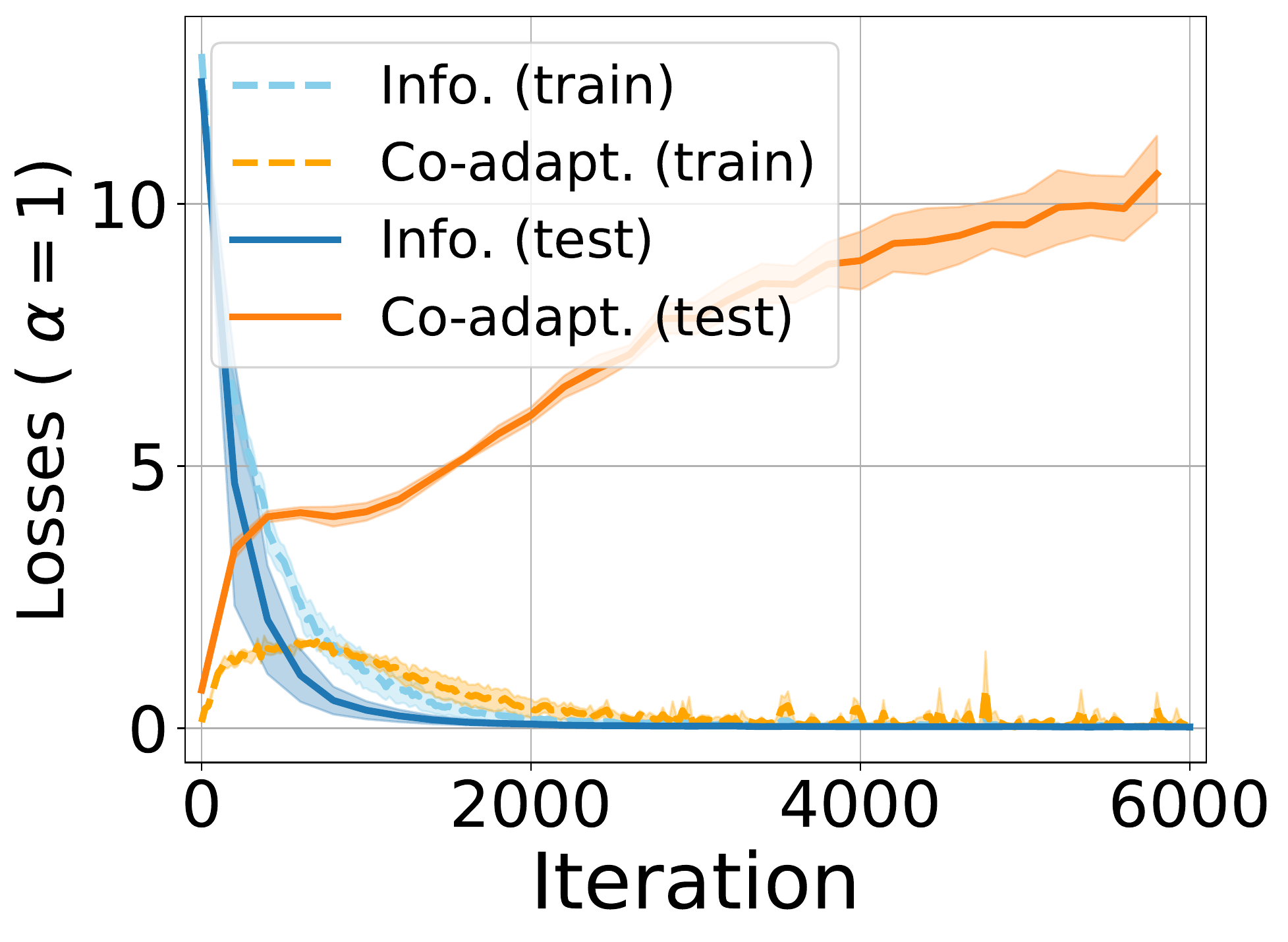}
\label{fig:1_dyn} 
\end{subfigure}
\hfill
\begin{subfigure}[t]{0.32\textwidth}
\centering
\includegraphics[width=\textwidth]{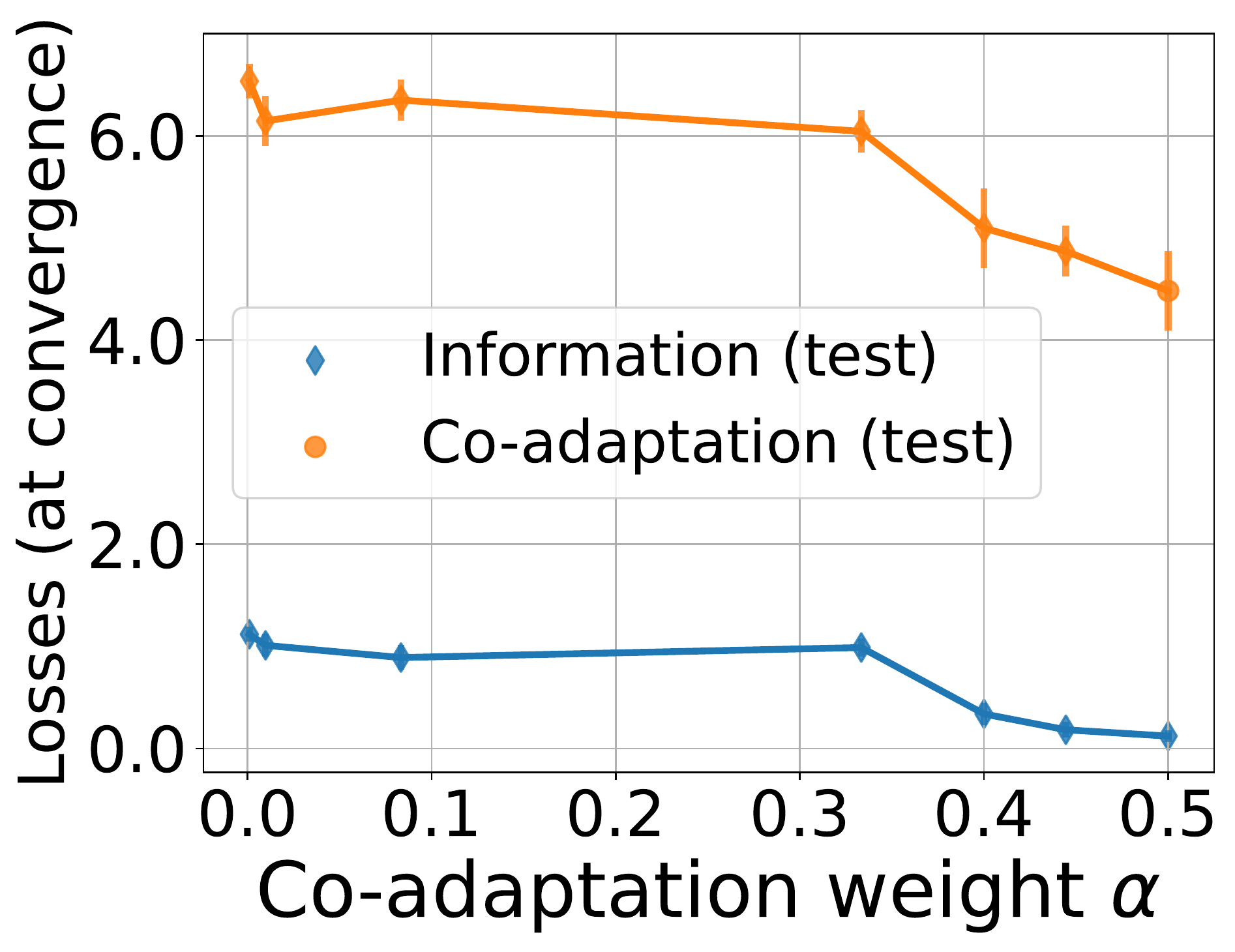}
\label{fig:1_losses} 
\end{subfigure}
\hfill
\begin{subfigure}[t]{0.33\textwidth}
\centering
\includegraphics[width=\textwidth]{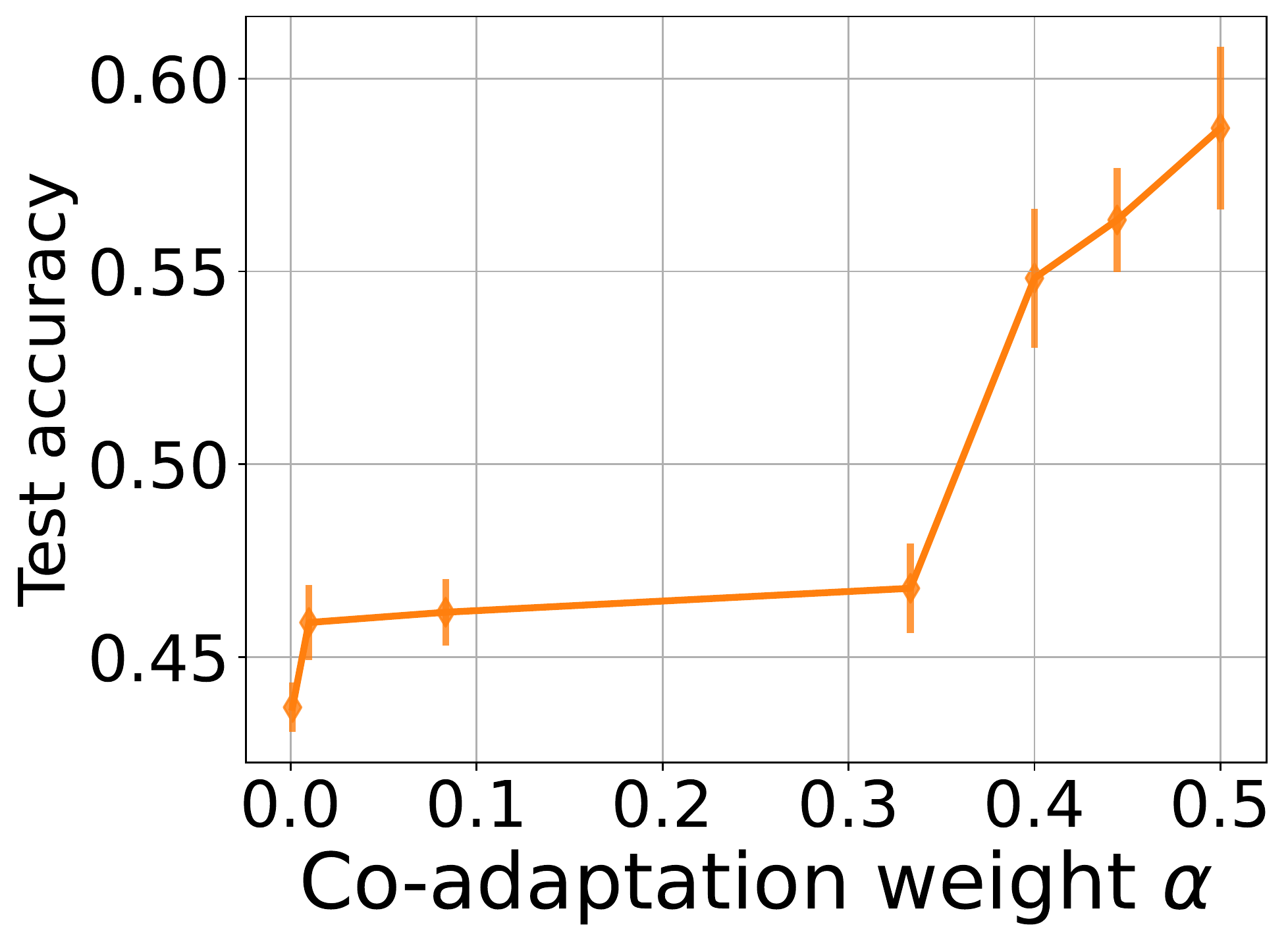}
\label{fig:1_gen} 
\end{subfigure}
\hfill
\vskip -1em
\caption{(a)Training dynamics ($\alpha=0.5$). (b,c)Agents score as a function of co-adaptation weight $\alpha$.
 \label{fig:alpha}}
\vskip -1.5em
\end{figure*} 

We here visualize the loss decomposition dynamics. Following the protocol of \ref{sec:task_balance_method}, we control $\Ladapt^{\mathrm{train}}$ in speaker's loss with weight $\alpha$ to understand the influence of the co-adaptation term on the language.


\label{sec:task_balance_expe}

\paragraph{The co-adaptation task overfits rapidly} We plot information and co-adaptation training dynamics in the standard setting ($\alpha=0.5$). Note that both train and test information losses quickly converge to $0$, in other words the speaker succeeds in developing a protocol that is unambiguous on both the training set and the overall distribution. On the other hand, the test co-adaptation loss diverges while the train co-adaptation keeps disminishing, highlighting a clear overfitting problem. 
%



\paragraph{The co-adaptation task promotes generalization}
We then display in Figure~\ref{fig:alpha} the evolution of the information and co-adaptation losses for different co-adaptation weight $\alpha$. We observe that down-weighting $\Ladapt^{\mathrm{train}}$ tends to enforce both information and co-adaptation overfitting. Thus, even though the co-adaptation loss is not inherently necessary for the speaker to develop an unambiguous language, it is important to encourage the speaker to build a better language. This is confirmed when looking at generalization accuracies. From $\alpha=0$ to $\alpha=0.5$, there is a gain of $15$ points of generalization. In conclusion, we note that (i) balancing the loss in favor of $\Linfo^{\mathrm{train}}$ has a negative impact on generalization, (ii) the co-adaptation loss $\Ladapt^{\mathrm{train}}$ pushes the speaker to develop a language that generalizes better. 


These experiments highlight two key findings: (i) co-adaptation is crucial for generalization ; (ii) in standard settings, the co-adaptation loss overfits substantially, whereas the information loss does not. 

\vspace{-0.25em}
\subsection{Countering co-adaptation overfitting}
\label{sec:expe_2}
\vspace{-0.25em}

We here investigate whether limiting overfitting in the co-adaptation loss may push towards languages that generalize better and are more stuctured. As described in \ref{sec:method_listener_cv_rate}, we compare three control baselines: \emph{Continuous listener}, \emph{Partial listener} with varying levels of convergence, and \emph{Early stopping listener}.

\begin{figure*}[th!]
\centering
\includegraphics[clip,trim=0.cm 3.9cm 0.cm 2.86cm, width=1.00\textwidth]{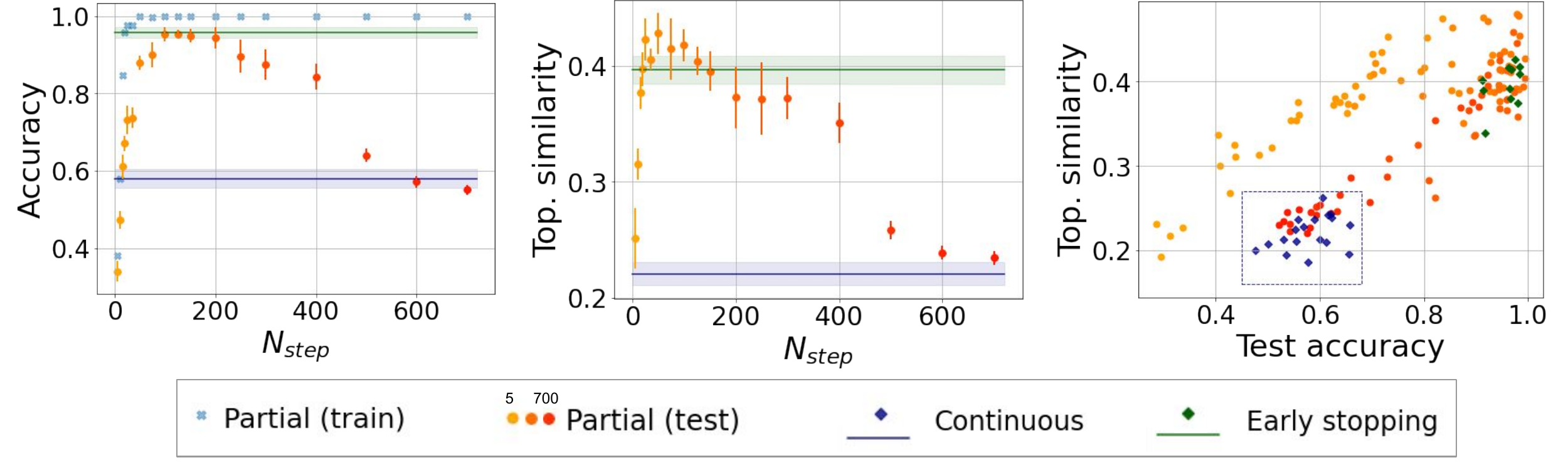}
\caption{(a,b) Evolution of generalization and top.sim with \emph{Partial listener}'s number of learning steps $N_{step}$ ; (c) Top. sim VS. generalization. The color level of orange dots increases with $N_{step}$. Blue (resp. green) lines and points refer to the Continuous listener (resp. Early stopping listener).}
\label{fig:evolution_generalization}
\end{figure*} 

\paragraph{Countering co-adaptation overfitting improves generalization}
In Figure~\ref{fig:evolution_generalization}, we observe that the level of convergence of the \emph{Partial listener} between each speaker's update (controlled by $N_{step}$) has a strong impact on the generalization of the emergent protocol. Overall, we recover classic machine learning trends when varying $N_{step}$: when $N_{step} < 50$, both train and test accuracy are  low --- the agents underfit. When $50 < N_{step} < 250$, the train and test accuracy are almost optimal --- the agents are in good training regime. Finally, when $N_{step} > 250$, the train accuracy is maximal while the test accuracy collapses --- the agents overfit. 
These observations reveal that the level of convergence of the listener has a substantial impact on the final emergent language capacity to generalize. Recall that, in these experiments, the direct effect of the listener's overfitting is mitigated, as we measure generalization using an auxiliary listener that is early stopped, and should therefore not overfit as noted in Section \ref{sec:exp_set}. The listener’s overfitting impacts the speaker’s update through the co-adaptation loss, which, by inducing a poorer final language leads to a degradation in generalization. Additionally, Figure~\ref{fig:evolution_generalization} shows that the continuous listener, standard in the Lewis games literature, provides generalization performance similar to the worst overfitting listeners.

Controlling the listener's co-adaptation level appears crucial to let the speaker develop a language that generalizes well; this effect may have been underestimated in the standard Lewis learning dynamic.

\paragraph{Countering co-adaptation overfitting improves compositionality}
Figure~\ref{fig:evolution_generalization} reveals that compositionality follows the same pattern. In the underfitting regime, the topographic similarity is low but still outperforms the \textit{Continuous listener}. Similarly, it is also low in the overfitting regime. In-between the two --- which corresponds to high generalization in Figure \ref{fig:evolution_generalization} --- the topographic similarity reaches high values, which suggests that more compositional languages emerge.
This indicates that the listener's lack of co-adaptation overfitting promotes structured languages.

\paragraph{Compositionality correlates with generalization}
In Figure~\ref{fig:evolution_generalization}, we plot the correlation between generalization and compositionality. As opposed to~\citep{chaabouni2020compositionality}, we observe a strong correlation between generalization and topographic similarity when varying the \emph{Partial listener}'s level of convergence. In particular, we identify two correlation branches: one belonging to the underfitting regime and the second to the overfitting regime. Together, they retrace the evolution of generalization and compositionality with respect to $N_{step}$. We see that \textit{Continuous listeners} belong to the end of this trajectory, in the overfitting regime. Note that the blue rectangle --- which delineates the range of values reached with the \textit{Continuous listener} --- corresponds to the classic learning setting in the literature. As this range is tight, it may explain the initial negative results reported by~\citep{chaabouni2020compositionality}.  

In conclusion, the listener exerts a necessary pressure on the speaker to develop a structured language that generalizes better. This pressure can be controlled by limiting the listener's level of overfitting, which is inevitably too high when the listener is trained continuously as is usually done. 

\paragraph{Comparison with standard regularization methods} 
\label{sec:compa_reg}
In practice, re-initializing the listener as done with the \emph{Partial} or \emph{Early stopping listener} is costly. We thus test whether performances comparable to Figure~\ref{fig:evolution_generalization} can be obtained by controlling the listener's level of overfitting with standard regularization methods. In Table~\ref{tab:reg_listener}, we report the influence of applying common regularization methods to the listener on various metrics of the language. We find that regularization consistently results in noticeable improvements. Moreover, once again, gains of generalization correlate with gains of compositionality. These trends corroborate our hypothesis that controlling the listener's learning is key to encourage the speaker to develop more structured languages. However, those methods remain under the upper bound reached by the \emph{Early stopping listener}, which suggests that further research on regularization in cooperative games is warranted. 

We complement this analysis in Appendix~\ref{app:reg_speaker} by studying the impact of regularization on the speaker's side, and show that such regularization does not result in similar improvements. This indicates that the listener is the main contributor to the co-adaptation overfitting. 

\vspace{-0.25em}
\subsection{Scaling to the Image Discrimination Games}
\vspace{-0.25em}
\label{sec:scaling_to_image}

\begin{table}[t]
\begin{adjustwidth}{-0.5cm}{-0.5cm}
\begin{center}
    \begin{minipage}[t][][b]{.5\linewidth}
    \scriptsize
    \centering
    \begin{tabular}{lccc}
    \toprule
     &  Gen. $\uparrow$ & Compo. $\uparrow$ & $\Ladaptestimate^{\mathrm{test}}$ $\downarrow$  \\ 
     \midrule \midrule
         Continuous & $0.58_{\pm0.05}$ & $0.22_{\pm0.02}$ & $4.64_{\pm1.22}$ \\
         \midrule
         Dropout & $0.64_{\pm0.03}$ & $0.24_{\pm0.01}$ & $4.86_{\pm0.52}$ \\
         No LN. & $0.70_{\pm0.03}$ & $0.24_{\pm0.02}$ & $4.68_{\pm0.38}$ \\
         Weight decay &  $0.72_{\pm0.03}$ & $0.25_{\pm0.03}$ & $4.29_{\pm0.56}$ \\
         No LN. + WD & $0.87_{\pm0.07}$ & $0.30_{\pm0.03}$ & $2.12_{\pm0.67}$ \\
         \midrule 
         Early stopping & $\mathbf{0.95_{\pm0.04}}$ & $0.39_{\pm0.04}$ & $1.10_{\pm0.69}$ \\
         Top Partial & $\mathbf{0.95_{\pm0.03}}$ & $\mathbf{0.42_{\pm0.02}}$ & $\mathbf{0.97_{\pm0.55}}$ \\
     \bottomrule
    \end{tabular}
    \end{minipage}%
    \hfill
    \begin{minipage}[t][][b]{.5\linewidth}
    \scriptsize
      \centering
        \begin{tabular}{lccc}
        \toprule
          \multicolumn{1}{l}{} & \multicolumn{2}{c}{Generalization $\uparrow$} \\ 
          \midrule
          \midrule
          \textbf{CelebA}     & 1/1 & $1/20$ & $1/100$   \\
          \midrule
          Continuous    & $0.94_{\pm0.01   }$ & $0.67_{\pm0.02}$          & $0.39_{\pm0.07}$  \\ 
          Early stopping    & $\mathbf{0.97_{\pm0.01   }}$ & $\mathbf{0.80_{\pm0.03}}$ & $\mathbf{0.69_{\pm0.04}}$  \\
          \midrule
          \midrule
          \textbf{ImageNet}      & 1/1 & $1/20$ & $1/100$   \\
          \midrule
          Continuous & $0.96_{\pm0.01   }$ & $0.77_{\pm0.01   }$ & $0.51_{\pm0.03}$  \\ 
          Early stopping    & $\mathbf{0.98_{\pm0.01   }}$ & $\mathbf{0.81_{\pm0.01}}$ & $\mathbf{0.64_{\pm0.01}}$  \\
        \bottomrule
        \end{tabular}
    \end{minipage} 
        \end{center}
        \end{adjustwidth}
    \vskip 1em
    \caption{(left) Performance comparisons between Continuous listener, Partial listener, Early stopping listener and classic listener regularization, e.g. weight decay~\cite{hinton1987learning, krogh1991simple}, Dropout~\cite{JMLR:v15:srivastava14a} and layernorm~\citep{ba2016layer}. Regularization parameters were tuned and are detailed in Appendix~\ref{app:reg_params} ; (right) Generalization scores for continuous baselines and Early stopping listener on visual Lewis Games. $1/1$, $1/20$ and $1/100$} refer to the subset ratios of the dataset.
    \label{tab:reg_listener}
\end{table}



\label{sec:exp_images}
To validate our empirical findings beyond synthetic games, we scale our approach to complex games with natural images as advocated by~\citep{chaabouni2022emergent}. We thus train our agents on a discriminative game on top of the CelebA~\citep{liu2015deep} and ImageNet~\citep{russakovsky2015imagenet,deng2009imagenet} datasets while applying previous protocol. 
We work on 3 sizes of training set with increasing  generalization difficulty.
We provide all the training details and game settings in Appendix~\ref{app:image_expe} and report our results in Table~\ref{tab:reg_listener}.
While agents generalize well when trained on the entire training set, generalization issues occur on smaller training sets and performances can indeed be improved by controlling the listener's level of convergence.
However, Appendix~\ref{app:top_sim_images} shows that gain of generalization does not correlate with gain of topographic similarity, supporting that agents' language structure is not captured by the topographic similarity in image based settings~\citep{chaabouni2022emergent,andreas2019measuring}.

\vspace{-0.5em}
\section{Related work}
\label{sec:related}
\vspace{-0.5em}

The decomposition of the loss function in the Lewis Game that we introduced finds echos in the cognitive science literature. 
According to Skyrms \citep{skyrms2010signals}, communicative organisms or systems are confronted with two types of information: about the environmental states shared by the agents (called \textit{objective} information), and about how an agent would react to a signal (called \textit{subjective} information). Communication protocols emerge as a trade-off between constraints related to those two types of information~\citep{kirby2002emergence,kirby2015compression}: the sender should be expressive~\citep{galantucci2011experimental,fay2013cultural} and transcribe the information available in the world with as little ambiguity as possible, which has been described as a \textit{bias against ambiguity}~\citep{spike2017minimal} ; sender and receiver should agree on the same referring system, which has been described as a \textit{conceptual pact}~\citep{brennan1996conceptual}. The latter has been shown to impose compressibility and learnability pressures promoting structure~\citep{tamariz2015culture,smith2003iterated,zaslavsky2018}. This analysis resonates well with our analytical decomposition of the loss function in the Lewis game.

The first term of the decomposition, which we called the information loss, has been addressed by  previous work that assumed that linguistic structure and generalization emerge from the requirement of creating an unambiguous language. In this line of work,  studies have either manipulated the complexity of the environment~\citep{chaabouni2022emergent,guo2021expressivity,slowik2002exploring,mu2021emergent}, restricted the bandwidth of the communication channel~\citep{kottur-etal-2017-natural,resnick2019capacity}, or added noise to the message~\citep{kucinski2021catalytic,Kuciski2020EmergenceOC}. In our main experiment, we do not apply such information constraints to better focus on the second term of the decomposition, the \textit{co-adaptation} constraint, less studied within a machine learning approach. Previous work have assumed that the co-adaptive dynamics encourage speakers to develop a more structured language for learnability reasons~\citep{li2019ease}. Support for this hypothesis can be found directly via the implementation of a neural variant of Iterated Learning~\citep{ren2020compositional} or the introduction of learning speed heterogeneities~\citep{rita2022on} and indirectly via the restriction of agents capacity~\citep{resnick2019capacity}, the variation of the communication-graph in populations~\citep{harding-graesser-etal-2019-emergent,kim2021emergent} or the addition of newborn agents~\citep{cogswell2020emergence}. In our paper, we demonstrate  that a co-adaptation term is always present in standard agents optimization protocols and show that controlling \textit{co-adaptation overfitting} enhances language properties. The existence of an overfitting regime found under the default setting (continuous training) may explain the counter-intuitive lack of relationship between compositionality and generalization previously reported with neural agents~\citep{lazaridou2020emergent,chaabouni2020compositionality,kharitonov-baroni-2020-emergent,dessi2019cnns}.



\vspace{-0.75em}
\section{Conclusion}
\vspace{-0.75em}


In this paper, we propose a methodological approach to better understand the dynamics in Lewis signaling games for language emergence. It allows us to surface two components of the training: (i) an information loss, (ii) a co-adaptation loss. We shed light that the agents tend to overfit this co-adaptation term during training, which hinders the learning dynamic and degrades the resulting language. As soon as this overfitting is controlled, agents develop compositional languages that better generalize. Remarkably, this emergent compositionality does not result from environmental factors, e.g. communication bottleneck~\citep{kirby2001spontaneous}, under-parametrization~\citep{kottur-etal-2017-natural,galke2022emergent}, population dynamics~\citep{chaabouni2022emergent,rita2022on}, memory restriction~\citep{cogswell2020emergence,cornish2017sequence} or inductive biases~\citep{rita-etal-2020-lazimpa}, but only through a trial-and-error process. Therefore, we advocate for a better comprehension of the optimization and machine learning issues. As illustrated in this paper, such understanding may unveil contradictions between computational models and language empirical observations and better expose the existing synergies between learning dynamics and environmental factors~\citep{Lupyan2010,wray2007543,raviv2019compo,clyne1992linguistic,cultural2022learning,ellis2008dynamics}.

\section*{Acknowledgments}

Authors would like to thank Rahma Chaabouni, Marco Baroni, Paul Smolensky, Bilal Piot and Karl Tuyls for helpful discussions and the anonymous reviewers to their relevant comments. M.R would also like to thank Michael Sander and Maureen de Seyssel for last minute feedbacks. M.R. was supported by the MSR-Inria joint lab and granted access to the HPC resources of IDRIS under the allocation 2021-AD011012278 made by GENCI. P.M. was supported by the ENS-CFM Data Science Chair. E.D. was funded in his EHESS role by the European Research Council (ERC-2011-AdG-295810 BOOTPHON), the Agence Nationale pour la Recherche (ANR-17-EURE-0017 Frontcog, ANR-10-IDEX0001-02 PSL*, ANR-19-P3IA-0001 PRAIRIE 3IA Institute) and grants from CIFAR (Learning in Machines and Brains) and Meta AI Research (Research Grant).

\clearpage


{
\small
\bibliography{biblio}
\bibliographystyle{plainnat}
}














\clearpage

\section*{Checklist}


\begin{enumerate}

\item For all authors...
\begin{enumerate}
  \item Do the main claims made in the abstract and introduction accurately reflect the paper's contributions and scope?
    \answerYes{}
  \item Did you describe the limitations of your work?
    \answerYes{We mentioned the Partial and Early Stopping listener training are costly opening a discussion on regularization methods (Section~\ref{sec:expe_2})}
  \item Did you discuss any potential negative societal impacts of your work?
    \answerNo{We run experiments that have no negative societal impacts}
  \item Have you read the ethics review guidelines and ensured that your paper conforms to them?
    \answerYes{}
\end{enumerate}

\item If you are including theoretical results...
\begin{enumerate}
  \item Did you state the full set of assumptions of all theoretical results?
    \answerYes{We took a large part of our analytical part to detail the set of assumptions. Further discussions to more general cases are discussed in Appendix~\ref{app:sec:lewis_demo}}
        \item Did you include complete proofs of all theoretical results?
    \answerYes{If not in the main paper, all derivations are brought in the Appendices}
\end{enumerate}

\item If you ran experiments...
\begin{enumerate}
  \item Did you include the code, data, and instructions needed to reproduce the main experimental results (either in the supplemental material or as a URL)?
    \answerNo{Code will be released upon deanonymization.}
  \item Did you specify all the training details (e.g., data splits, hyperparameters, how they were chosen)?
    \answerYes{}
        \item Did you report error bars (e.g., with respect to the random seed after running experiments multiple times)?
    \answerYes{}
        \item Did you include the total amount of compute and the type of resources used (e.g., type of GPUs, internal cluster, or cloud provider)?
    \answerNo{Our work was carried out on GPUs on an institutional cluster. It is will be mentionned upon deanonymization. Each experiment runs on a single V100-32G GPU}
\end{enumerate}

\item If you are using existing assets (e.g., code, data, models) or curating/releasing new assets...
\begin{enumerate}
  \item If your work uses existing assets, did you cite the creators?
    \answerYes{The ImageNet and CelebA papers were cited}
  \item Did you mention the license of the assets?
    \answerNo{}
  \item Did you include any new assets either in the supplemental material or as a URL?
    \answerNA{}
  \item Did you discuss whether and how consent was obtained from people whose data you're using/curating?
    \answerNA{}
  \item Did you discuss whether the data you are using/curating contains personally identifiable information or offensive content?
    \answerNA{}
\end{enumerate}

\item If you used crowdsourcing or conducted research with human subjects...
\begin{enumerate}
  \item Did you include the full text of instructions given to participants and screenshots, if applicable?
    \answerNA{}
  \item Did you describe any potential participant risks, with links to Institutional Review Board (IRB) approvals, if applicable?
    \answerNA{}
  \item Did you include the estimated hourly wage paid to participants and the total amount spent on participant compensation?
    \answerNA{}
\end{enumerate}

\end{enumerate}


\clearpage 
\input{appendix}

\end{document}

%% file: appendix.tex
\appendix




\section{Lewis Games Loss Decomposition : proofs}
\label{app:sec:lewis_demo}

We provide all the proofs of the Lewis Games Loss Decomposition. We organize the proofs as follows:
\begin{itemize}
    \item \textbf{Appendix~\ref{app:sec:lewis_reco} - Reconstruction game}, we provide the proofs of the Decomposition for the reconstruction game. 
    \begin{itemize}
        \item \textbf{Appendix~\ref{app:sec:lewis_reco_log_reward} - log-likelihood reward}, we first prove the loss decomposition when the reward is the reconstruction log-likelihood (case of the main paper).
        \item \textbf{Appendix~\ref{app:sec:lewis_reco_gen_reward} - general reward}, we then extend the decomposition to a more general reward
    \end{itemize}
    \item \textbf{Appendix~\ref{app:sec:general_proof} - Extension to Lewis games}, we extend the Loss Decomposition to a more general class of Lewis games. We first describe the additional formalism (Appendix~\ref{app:sec:formalism}), then we prove the decomposition when the reward is the listener's log-likelihood (Appendix~\ref{app:sec:proof_gen_log}) and when the reward is more general (Appendix~\ref{app:sec:lewis_gen_gen_reward}). Eventually, we show how the classic discrimination game can be expressed under this formalism in Appendix~\ref{app:sec:discri}.
    \item \textbf{Appendix~\ref{app:sec:asym_reward} - Extension to agents optimizing different rewards}, we discuss how the decomposition is affected when the agents optimize different rewards.
\end{itemize}


\subsection{Proof of the Lewis Reconstruction Game Loss Decompositon}
\label{app:sec:lewis_reco}

Let's first recall some notations that we will use throughout the proofs. We consider two agents: a speaker parameterized by $\theta$ and a listener parameterized by $\phi$. In the reconstruction game, the speaker observes objects denoted by $x$ and taken from a set $\mathcal{X}$. The random variable characterizing the object is denoted by $X$ and its distribution is denoted by $p$. Based on object $x$, the speaker then sends a message $m$ from a message space $\mathcal{M}$ according to its policy $\pi_{\theta}(.|x)$. The random variable $M_{\theta}$ characterizes the message that is sampled from the speaker's policy $\pi_{\theta}$. Eventually, the listener should reconstruct the original object $x$ based on the message $m$. The probability that the listener predicts the input $x$ given a message $m$ is denoted by $\rho_{\phi}(x|m)$. 

For any probability distribution, we denote by $\mathrm{Supp}$ the support of the distribution.

In the reconstruction game, the two agents optimize the same loss:

\begin{align*}
        \mathcal{L}_{\theta,\phi} &= - \mathbb{E}_{x \sim p, m \sim \speaker(\cdot|x)}[r_{\phi}(x,m)]
\end{align*}

We will first prove the decomposition in the case where $r_{\phi}(x,m)=\log \listener(x|m)$ (reconstruction log-likelihood) for all $x$ and $m$ and then for a more general form of reward.

\subsubsection{Proof of the Decomposition when $r_{\phi}(x,m)=\log \rho_{\phi}(x|m)$}
\label{app:sec:lewis_reco_log_reward}

We first prove the decomposition in the case described in the main paper: $r_{\phi}(x,m)=\log\listener(x|m)$ for all $x$ and $m$.

\paragraph{Optimal listener} For completeness, we recall the proof of Equation~\ref{inverse_bayes} of the expression of the listener that is optimal with respect to $\mathcal{L}_{\theta,\phi}$.

In the case $r_{\phi}(x,m)=\log \rho_{\phi}(x|m)$, the listener is optimizing a cross-entropy loss with respect to the joint variable $(X,M)$ where $X$ follows $p$ and $M$ follows speaker's policy $\pi_{\theta}$. The loss can be rewritten as:

\begin{align*}
    \mathcal{L}_{\theta,\phi} & = -\mathbb{E}_{x \sim p, m \sim \pi_{\theta}(.|x)}[\log\rho_{\phi}(x|m)] \\
    \mathcal{L}_{\theta,\phi} & = -\mathbb{E}_{m \sim \pi_{\theta}, x \sim \listeneropt(.|m)}[\log\rho_{\phi}(x|m)]
\end{align*}

According to Gibbs inequality, the optimal distribution $\rho_{\phi^{*}}(\cdot|m)$ for all $m$ is $\rho_{\phi^{*}}(.|m)=\listeneropt(.|m)$ where $\listeneropt(.|m)$ is the speaker's posterior distribution with respect to the prior $p$ and the conditional distribution $\pi_{\theta}(\cdot|x)$:

\begin{center}
\begin{align*}
    \listeneropt(x|m):=\frac{p(x)\pi_{\theta}(m|x)}{\mathbb{E}_{x' \sim p}[p(x')\pi_{\theta}(m|x')]} \quad \quad \text{for all} \quad  x \in \mathrm{Supp}(p), m \in \mathrm{Supp}(\pi_{\theta}(\cdot|x))
\end{align*}
\end{center}

This concludes the proof of Equation~\ref{inverse_bayes}. 

\paragraph{Loss Decomposition} The idea of the proof is to decompose the reward into the optimal reward (when the listener is optimal), denoted by $r^{*(\theta)}(x,m)$, and the residual that measures the optimality gap, denoted by $r_{\phi}(x,m)-r^{\theta}(x,m)$:

\begin{align*}
    r_{\phi}(x,m)= r^{*(\theta)}(x,m) + (r_{\phi}(x,m)-r^{*(\theta)}(x,m)) \qquad \text{for all} \qquad x \in \mathrm{Supp}(p) , m \in \mathrm{Supp}(\pi_{\theta}(.|x))
\end{align*}

Due to the linearity of the expectation, it follows that:

\begin{align*}
    \mathcal{L}_{\theta,\phi}= -\mathbb{E}_{x \sim p, m \sim \pi_{\theta}(.|x)}[r^{*(\theta)}(x,m)] - \mathbb{E}_{x \sim p, m \sim \pi_{\theta}(.|x)}[r_{\phi}(x,m)-r^{*(\theta)}(x,m)]
\end{align*}

In the case where the reward is taken as the listener's log-likelihood, we have: 

\begin{align*}
    \mathcal{L}_{\theta,\phi} &= -\mathbb{E}_{x \sim p, m \sim \pi_{\theta}(.|x)}[r^{*(\theta)}(x,m)] - \mathbb{E}_{x \sim p, m \sim \pi_{\theta}(.|x)}[r_{\phi}(x,m)-r^{*(\theta)}(x,m)] \\ 
    &= -\mathbb{E}_{x \sim p, m \sim \pi_{\theta}(.|x)}[\log \listeneropt(x|m)] - \mathbb{E}_{x \sim p, m \sim \pi_{\theta}(.|x)}\left[\log\frac{\rho_{\phi}(x|m)}{\listeneropt(x|m)}\right] \\ 
    &= -\mathbb{E}_{x \sim p, m \sim \pi_{\theta}(.|x)}[\log \listeneropt(x|m)] - \mathbb{E}_{m \sim \pi_{\theta}}\mathbb{E}_{x\sim \listeneropt(.|m)}\left[\log\frac{\rho_{\phi}(x|m)}{\listeneropt(x|m)}\right]  \\
   \mathcal{L}_{\theta,\phi} &=   \underbrace{\mathcal{H}(X|M_{\theta})}_{\Linfo} + \underbrace{\mathbb{E}_{m\sim \speaker}D_{KL}(\listeneropt(\cdot|m)||\listener(\cdot|m))}_{\Ladapt}
\end{align*}

where $\mathcal{H}(X|M_{\theta})$ is the conditional entropy of $X$ conditioned on $M_{\theta}$ and $D_{KL}(p||q)$ is the Kullback-Leiber divergence between two distributions $p$ and $q$.

This last computation concludes the proof of Equation~\ref{eq:SGTD}.

\paragraph{Remarks} The key ingredients of the loss decomposition are:
\begin{enumerate}
    \item We isolate two sub losses: $\Linfo$, independent from the listener ; $\Ladapt$ optimized both by the speaker and the listener.
    \item $\Linfo$ measures the degree of ambiguity in the communication protocol. If $\Linfo$ is optimal, ie. $\Linfo = 0$, messages are unambiguous: each message refers to a unique input. Otherwise, $\Linfo>0$ and ambiguities remain.
    \item $\Ladapt$ measures the gap between the listener and its optimum (here the speaker's posterior distribution). When the listener is optimal, $\Ladapt=0$ and the main loss is limited to its information part, otherwise $\Ladapt > 0$ and the speaker and listener should adapt to reduce the optimality gap.
\end{enumerate}

\subsubsection{Decomposition with a General Reward}
\label{app:sec:lewis_reco_gen_reward}

In order to generalize the loss decomposition to more general rewards, we adopt the following strategy:

\begin{itemize}
    \item \textbf{Construction of the reward}: we first need to build a general expression of the communication reward. To do so, we describe the conditions that the cooperative reward should fulfill in the reconstruction game and then propose a general reward expression. For the sake of generality, we consider that the environment $\mathcal{X}$ and message space $\mathcal{M}$ may be continuous spaces and that all the probability distribution may not be discrete.
    \item \textbf{Examples of usual cases}: we show that our proposed general expression covers the rewards used in most emergent communication papers, e.g. log-likelihood and accuracy.
    \item \textbf{Loss decomposition in the general case}: we write the loss decomposition with this general form of reward, showing that the key properties of the loss decomposition still hold.
\end{itemize}

\paragraph{Construction of the reward}
The Lewis reconstruction game is a cooperative game: the more the listener is able to reconstruct the objects seen by the speaker, the better the task is solved both by the speaker and the listener. Therefore the reward of the Lewis reconstruction game should respect the following conditions:
\begin{itemize}
    \item \textbf{C1}: For $x \in \mathrm{Supp}(p)$ and $m \in \mathrm{Supp}(\pi_{\theta}(.|x))$, the expected reward $r_{\phi}(x,m)$ is maximum when $\rho_{\phi}(.|m)=\mathbf{1}_{x}$, where ie. $\mathbf{1}_{x}$ denotes the indicator function on $\mathcal{X}$ taken on $x$: the listener predicts $x$ with probability $1$ when it receives $m$.
    \item \textbf{C2}: For $x \in \mathrm{Supp}(p)$ and $m \in \mathrm{Supp}(\pi_{\theta}(.|x))$, the expected reward $r_{\phi}(x,m)$ is sub-optimal when $\rho_{\phi}(.|m) \ne \mathbf{1}_{x}$, ie., the listener has a non-negative probability to predict the wrong object $x' \ne x$.
\end{itemize}

Given these assumptions, we propose the following general reward expression:

\begin{align}
    r_{\phi}(x,m) = - D(\mathbf{1}_{x}||\rho_{\phi}(.|m)) + K
    \label{eq:general_reward}
\end{align}

where $D$ is such that $D(p||q) = 0$ iff  $p=q$,$D(p||q) > 0$ otherwise. $\mathbf{1}_{x}$ is the indicator function on $\mathcal{X}$ taken in $x$  and $K$ is a real number that fixes the highest value of the reward. Note that $D(p||q)$ is close to a divergence, but has less assumptions.

\paragraph{Usual rewards as special instances of the general expression}

We show that Equation~\ref{eq:general_reward} recovers most rewards used in the emergent communication literature, specifically:

\begin{itemize}
    \item \textbf{Reconstruction log-likelihood}~\citep{chaabouni2019anti, chaabouni2020compositionality, kharitonov2020entropy, rita2022on, rita-etal-2020-lazimpa, chaabouni2021communicating} This case is the one used in the main paper and which is standarly used in reconstruction settings. With the following parameters
    \begin{itemize}
        \item $D(p||q)=D_{KL}(p||q),$
        \item $K=0\cdot$
    \end{itemize}
    we have:
    \begin{align*}
        r_{\phi}(x,m)=- D_{KL}(\mathbf{1}_{x}||\rho_{\phi}(.|m))=\log \rho_{\phi}(x|m) 
    \end{align*}
    
    \item \textbf{Accuracy}~\citep{ren2020compositional, lazaridou2018emergence, lazaridou2016multi, kottur-etal-2017-natural, li2019ease, harding-graesser-etal-2019-emergent, evtimova2018emergent}  Accuracy is the most commonly used reward in the emergent communication literature. It corresponds to agents receiving reward $1$ if the prediction sampled according to the listener's output probability $\listener(\cdot|m)$ matches the original object $x$. The pointwise accuracy depends on the specific sample drawn from the listener's distribution. We are interested in how good the listener's prediction is on average, and thus in the expected accuracy, which is expressed as in Equation \ref{eq:general_reward} with:
    \begin{itemize}
        \item $D(p||q) = 1 - \mathbb{E}_{p}[q]$ 
        \item $K = 1$
    \end{itemize}
    
    The expected accuracy is then defined as:
    
    \begin{align*}
        r_{\phi}(x,m)= D(\mathbf{1}_{x}||\rho_{\phi}(.|m))=\rho_{\phi}(x|m)\cdot
    \end{align*}
    
    
\end{itemize}

\paragraph{Loss Decomposition}

With this definition of the reward, the speaker and listener loss can be written:

\begin{align*}
    \mathcal{L}_{\theta,\phi}&=-\mathbb{E}_{x\sim p, m \sim \speaker(\cdot|x)}[r_{\phi}(x|m)] \\ 
    &= -K + \mathbb{E}_{x\sim p, m \sim \speaker(\cdot|x)}[D(\mathbf{1}_{x}||\rho_{\phi}(.|m))]
\end{align*}

We first need to define the optimal listener. Note that the expectation can be re-formulated:

\begin{align*}
    \mathcal{L}_{\theta,\phi} &= -K + \mathbb{E}_{m\sim \pi_{\theta}, x \sim \listeneropt(\cdot|m)}[D(\mathbf{1}_{x}||\rho_{\phi}(.|m))]
\end{align*}

The optimal listener is the listener $\rho_{\phi}$ that minimizes $\mathbb{E}_{x \sim \listeneropt(\cdot|m)}[D(\mathbf{1}_{x},\rho_{\phi}(.|m))]$ for all $m$. In the general case, there is no close-formed expression of the optimal listener. The optimal listener policy is dependent of the function $D$ and the posterior distribution $\listeneropt(\cdot|m)$. We next denote the optimal listener policy $\rho_{\phi^{*}}(\cdot|m)$ that is fully characterized by $\listeneropt(\cdot|m)$ and $D$ and is \textbf{independent} of $\phi$.


As in Appendix~\ref{app:sec:lewis_reco_log_reward}, we denote $r^{*(\theta)}(x,m)$ the reward of the optimal listener. We can then apply the same reward decomposition as in Appendix~\ref{app:sec:lewis_reco_log_reward}:

\begin{align*}
    r_{\phi}(x,m) &= r^{*(\theta)}(x,m) + (r_{\phi}(x,m)-r^{*(\theta)}(x,m)) \qquad \text{for all} \qquad x \in \mathrm{Supp}(p) , m \in \mathrm{Supp}(\pi_{\theta}(.|x))
\end{align*}

which is equal to:

\begin{align*}
    r_{\phi}(x,m) &= - D(\mathbf{1}_{x}||\rho_{\phi^{*}}(.|m)) -  [D(\mathbf{1}_{x}||\rho_{\phi}(.|m)) - D(\mathbf{1}_{x}||\rho_{\phi^{*}}(.|m))] - K
\end{align*}

where $\rho_{\phi^{*}}(.|m)$ is the optimal listener distribution that is independent of $\phi$.

We can then rewrite the loss by taking the expectation of this reward and isolate an information and co-adaptation component:

\begin{align*}
    \mathcal{L}_{\theta,\phi} &= -\mathbb{E}_{x \sim p, m \sim \pi_{\theta}(.|x)}[r^{*(\theta)}(x,m)] - \mathbb{E}_{x \sim p, m \sim \pi_{\theta}(.|x)}[r_{\phi}(x,m)-r^{*(\theta)}(x,m)] \\ 
    &= \underbrace{\mathbb{E}_{m \sim \speaker, x \sim \listeneropt(.|m)}[D(\mathbf{1}_{x}||\rho_{\phi^{*}}(\cdot|m))]}_{\Linfo} \\ 
    & + \underbrace{\mathbb{E}_{m \sim \speaker, x \sim \listeneropt(.|m)}[D(\mathbf{1}_{x}||\rho_{\phi}(.|m)) - D(\mathbf{1}_{x}||\rho_{\phi^{*}}(\cdot|m))]}_{\Ladapt}- K
\end{align*}

To be an information/co-adaptation decomposition, this loss decomposition should fulfill the following conditions:

\begin{enumerate}
    \item $\Linfo$ should be independent from the listener's weight $\phi$; $\Ladapt$ should be optimized both by the speaker and the listener.
    \item $\Linfo$ should be optimal ($\Linfo=0$) when the communication protocol is unambiguous, ie. each message refers to a unique input, sub-optimal ($\Linfo>0$) otherwise.
    \item $\Ladapt$ should be $0$ when the listener matches its optimum value with respect to the current object-message joint distribution, otherwise $\Linfo>0$ and the speaker and listener should adapt to reduce the optimality gap.
\end{enumerate}

Let's prove that all those conditions hold:

\begin{enumerate}
    \item The optimal listener policy $\rho_{\phi^{*}}(\cdot|m)$ is independent of $\phi$. It turns out that $\Linfo$ is independent from the listener. On the contrary, $\Ladapt$ is dependent both on $\theta$ and $\phi$ and therefore is optimized both by the speaker and listener.
    
    \item Let first show that when $\Linfo=0$, the speaker language is unambiguous. The language is considered unambiguous iff each message refers to a unique input. Formally, let $x$ be in the support of $p$ and 
 \begin{align*}
     \mathcal{M}_{x}=\{m \in \mathcal{M} \quad | \quad \listeneropt(x|m)>0\},
 \end{align*}
 be the set of messages referring to $x$. 
 
 This set is non empty because $\mathbb{E}_{m \sim \pi_{\theta}}[\listeneropt(x|m)]=p(x)>0$. The emergent language is considered unambiguous iff for all $x$ and $x'$ in the support of $p$: 
 \begin{align*}
     x\ne x' \Rightarrow \mathcal{M}_{x}\cap \mathcal{M}_{x'}=\emptyset,
 \end{align*}
 
 This property is equivalent of having a speaker posterior distribution $\listeneropt(.|m)$ being a Dirac distribution for all $m$ (otherwise, there is at least one message that refers to more than one object).
 
 Let's demonstrate that $\Linfo=0$ iff $\listeneropt(.|m)$ is a Dirac distribution for all $m$.
 
 First, when the speaker's posterior distribution is not a Dirac distribution, we have: $\Linfo>0$. 
 Let $m$ be a message in the support of $\pi_{\theta}$. If $\listeneropt(\cdot|m)$ is not a Dirac distribution, there exists $x$ such that $\listeneropt(x|m)>0$ and $D(\mathbf{1}_{x}||\rho_{\phi^{*}}(\cdot|m))>0$. Indeed, if there exists $x'$ such that $D(\mathbf{1}_{x'}||\rho_{\phi^{*}}(\cdot|m))=0$, we have $\rho_{\phi^{*}}(\cdot|m)=\mathbf{1}_{x'}$ by definition of $D$ and thus: if $x\ne x' \quad \Rightarrow \quad D(\mathbf{1}_{x}||\rho_{\phi^{*}}(\cdot|m))=D(\mathbf{1}_{x}||\mathbf{1}_{x'})>0$ by definition of $D$. It implies that when $\listeneropt(\cdot|m)$ is not a Dirac distribution : $\mathbb{E}_{x \sim \listeneropt(.|m)}[D(\mathbf{1}_{x}||\rho_{\phi^{*}}(\cdot|m))]>0$.
 
 Reciprocally, if for all $m \in \mathrm{Supp}(\speaker)$, $\listeneropt(\cdot|m)$ is a Dirac distribution: $\listeneropt(\cdot|m)=\mathbf{1}_{x_{m}}$ (with $m$ referring to $x_{m}$ and all $x \in \mathrm{Supp}(p)$ covered by the messages) and the corresponding optimal listener is also the Dirac distribution $\rho_{\phi^{*}}(\cdot|m)=\mathbf{1}_{x_{m}}$, we have:
 \begin{align*}
     \Linfo = \mathbb{E}_{m \sim \speaker}[D(\mathbf{1}_{x_{m}}||\rho_{\phi^{*}}(\cdot|m))]=\mathbb{E}_{m \sim \speaker}[D(\mathbf{1}_{x_{m}}||\mathbf{1}_{x_{m}})]=0
 \end{align*}
 
 Therefore, $\Linfo$ is equal to $0$, ie. is minimum, if and only if the speaker has a posterior which is Dirac distribution, ie. the speaker develops an unambiguous language.
    
    \item When the listener is optimal with respect to its loss, $\rho_{\phi}(.|m)=\rho_{\phi^{*}}(.|m)$ for all $m$ and as a direct consequence, $\Ladapt=0$. When the listener is not optimal with respect to its loss, $\Ladapt>0$ by definition of the optimal listener which is the listener that minimizes $\mathbb{E}_{m\sim \pi_{\theta}, x \sim \listeneropt(\cdot|m)}[D(\mathbf{1}_{x}||\rho_{\phi}(.|m))]$.
\end{enumerate}

In conclusion, in the case of a general reward, we keep the main ingredients of the information/co-adaptation decomposition.





\subsection{General Proof of the Lewis Games Loss Decomposition}
\label{app:sec:general_proof}

In the previous section, we provided a proof of the loss decomposition for the Lewis Reconstruction Game with a general cooperative reward. The goal of this Section is to extend this decomposition to a more general definition of Lewis Games:

\begin{itemize}
    \item \textbf{Appendix~\ref{app:sec:formalism} - Formalism}: We first describe the additional formalism.
    \item \textbf{Appendix~\ref{app:sec:proof_gen_log} - Log-likelihood reward}: We prove the decomposition for the general Lewis Game when the reward is the listener's log-likelihood.
    \item \textbf{Appendix~\ref{app:sec:discri} - General cooperative reward} We prove the decomposition for the general Lewis Game with a general cooperative reward.
    \item \textbf{Appendix~\ref{app:sec:discri} - Discrimination game} : Eventually, we show how the widely studied discrimination game~\citep{chaabouni2022emergent, mu2021emergent, dessi2021interpretable, guo2021expressivity, ren2020compositional, lazaridou2018emergence, lazaridou2016multi, havrylov2017emergence, li2019ease, Lowe2020On} can be expressed under this formalism.
\end{itemize}

\subsubsection{Formalism}
\label{app:sec:formalism}

In the general form, we consider inputs $x$ from a set $\mathcal{X}$ where $x$ is drawn from $p_{X}$. We consider a random feature $F$ of $X$ (in the reconstruction game $F=X$) that is distributed following $p_{F}(.|X)$. A draw of $F$ is denoted $f$ and the set of potential features $\mathcal{F}$. We here consider that the listener may have access to an auxiliary input $y$. We denote $Y$ the random variable of this auxiliary input and  $p_{Y}(\cdot|X,F)$ its probability distribution. The task is here the communication of the feature $f$. To this end, the speaker still sends messages $m$ from the message space $\mathcal{M}$. The random variable $M_{\theta}$ characterizes the messages that are sampled from the speaker's policy $\pi_{\theta}(\cdot|X)$. Eventually, the probability that the listener predicts the correct feature $f$, given message $m$ and auxiliary features $y$ is denoted by $\rho_{\phi}(f|m, y)$.

\subsubsection{Proof with the log-likelihood reward: $r_{\phi}(f,m,y)=\log \rho_{\phi}(f|m,y)$}
\label{app:sec:proof_gen_log}

We first prove the decomposition in the case: $r_{\phi}(f,m,y)=\log\listener(f|m,y)$ for all $f$,$m$ and $y$, ie. the reward is the listener's log-likelihood of predicting the good feature. The agents' loss becomes

\begin{align*}
    \mathcal{L}_{\theta,\phi} &= - \mathbb{E}_{x \sim p_{X},f \sim p_{F}(\cdot|x),y \sim p_{Y}(\cdot|x,f), m \sim \speaker(\cdot|x)}[\log\listener(f|m,y)]\cdot
\end{align*}


\paragraph{Optimal listener} The optimal listener is the listener that optimally minimizes $L_{\theta,\phi}$ for a fixed speaker policy $\pi_{\theta}$. It is obtained by noting that:

\begin{align*}
    \mathcal{L}_{\theta,\phi} &= - \mathbb{E}_{x \sim p_{X},f \sim p_{F}(\cdot|x),y \sim p_{Y}(\cdot|x,f), m \sim \speaker(\cdot|x)}[\log\listener(f|m,y)] \\
    \mathcal{L}_{\theta,\phi} &= - \mathbb{E}_{(m, y) \sim p_{M_\theta, Y}}\mathbb{E}_{f \sim \listeneropt(\cdot|m,y)}[\log\listener(f|m,y)] \cdot
\end{align*}
where $\listeneropt(f|m,y)=\frac{\mathbb{E}_{x \sim p_{X}}[ \pi_\theta(m|x) p_F(f|x) p_Y(y|f, x)]}{\mathbb{E}_{x \sim p_{X},f \sim p_{F}(.|x)}[\pi_\theta(m|x) p_Y(y|f, x)]}$ for all $f$, $m$ and $y$ and $p_{M_\theta, Y}(m, y) = \mathbb{E}_{x \sim p_{X},f \sim p_{F}(.|x)}[ \pi_\theta(m|x)p_Y(y|f, x)]$. 

It follows from Gibbs inequality that the optimal listener is $\listeneropt(\cdot|m,y)$ for all $m$ and $y$.

We can apply the reward decomposition of Appendix~\ref{app:sec:lewis_reco_log_reward}:

\begin{align*}
    r_{\phi}(x,m)= r^{*(\theta)}(f,m,y) + (r_{\phi}(f,m,y)-r^{*(\theta)}(f,m,y)) \qquad \text{for all} \qquad f \in \mathcal{F},m \in \mathcal{M},y \in \mathcal{Y}\cdot
\end{align*}

Plugging this decomposed reward in our loss, and applying the exact same steps as in Appendix~\ref{app:sec:lewis_reco_log_reward}, we get
\begin{equation}
   \mathcal{L}_{\theta,\phi} =   \underbrace{\mathcal{H}(F|M_{\theta}, Y)}_{\Linfo} + \underbrace{\mathbb{E}_{(m, y)\sim p_{M_\theta, Y}}D_{KL}(\listeneropt(\cdot|m, y)||\listener(\cdot|m, y))}_{\Ladapt}
   \label{eq:gen_decompo}
\end{equation}

which is the Loss Decomposition for a general game. 

\paragraph{Remarks} You note that the decomposition is close to the Loss Decomposition in the reconstruction case (Equation~\ref{eq:SGTD}). Indeed, since the listener should predict a given feature $F$, the information task is to build an unambiguous message protocol with respect to this feature and the optimal listener becomes the posterior distribution of the speaker with respect to this feature. The co-adaptation loss is once again a Kullback-Leiber distribution between the listener and the speaker's posterior. $\Linfo$ and $\Ladapt$ respects the conditions states in Appendix~\ref{app:sec:lewis_reco_log_reward}.

\subsubsection{Proof with the general reward $r_{\phi}(f,m,y) = - D(\mathbf{1}_{f}||\rho_{\phi}(.|m,y)) + K$}
\label{app:sec:lewis_gen_gen_reward}

To study the general case, we use the reward definition provided in Appendix~\ref{app:sec:lewis_reco_gen_reward}:

\begin{align*}
    r_{\phi}(f,m,y) = - D(\mathbf{1}_{f}||\rho_{\phi}(.|m,y)) + K
\end{align*}

where $D(p||q)$ is a function that is null when $p=q$, greater than $0$ otherwise, $\mathbf{1}_{f}$ the indicator function on $\mathcal{F}$ taken in $f$  and $K$ is a real number that fixes the highest value of the reward.

Agents' loss becomes:

\begin{align*}
    \mathcal{L}_{\theta,\phi}&=-\mathbb{E}_{x \sim p_{X},f \sim p_{F}(\cdot|x),y \sim p_{Y}(\cdot|x,f), m \sim \speaker(\cdot|x)}[r_{\phi}(f,m,y)] \\ 
    \mathcal{L}_{\theta,\phi} &= \mathbb{E}_{x \sim p_{X},f \sim p_{F}(\cdot|x),y \sim p_{Y}(\cdot|x,f), m \sim \speaker(\cdot|x)}[D(\mathbf{1}_{f}||\rho_{\phi}(.|m,y))] - K
\end{align*}

Denoting $\rho_{\phi^{*}}(.|m,y)$ the listener that optimally minimises $\mathcal{L}_{\theta,\phi}$ and $r^{\theta}(f,m,y)$ the reward of the optimal listener, the loss can be decomposed:

\begin{align*}
    \mathcal{L}_{\theta,\phi}&=-\mathbb{E}_{x \sim p_{X},f \sim p_{F}(\cdot|x),y \sim p_{Y}(\cdot|x,f), m \sim \speaker(\cdot|x)}[r^{*(\theta)}(f,m,y)] \\
    &-\mathbb{E}_{x \sim p_{X},f \sim p_{F}(\cdot|x),y \sim p_{Y}(\cdot|x,f), m \sim \speaker(\cdot|x)}[r_{\phi}(f,m,y)-r^{*(\theta)}(f,m,y)]  \\ 
    \mathcal{L}_{\theta,\phi} &= \underbrace{\mathbb{E}_{x \sim p_{X},f \sim p_{F}(\cdot|x),y \sim p_{Y}(\cdot|x,f), m \sim \speaker(\cdot|x)}[D(\mathbf{1}_{f}||\rho_{\phi^{*}}(.|m,y))]}_{\Linfo}\\ 
    &+ \underbrace{\mathbb{E}_{x \sim p_{X},f \sim p_{F}(\cdot|x),y \sim p_{Y}(\cdot|x,f), m \sim \speaker(\cdot|x)}[D(\mathbf{1}_{f}||\rho_{\phi}(.|m,y))-D(\mathbf{1}_{f}||\rho_{\phi^{*}}(.|m,y))]}_{\Ladapt}  - K 
\end{align*}

For the same arguments as in Appendix~\ref{app:sec:lewis_reco_gen_reward}, $\Linfo$ is only optimized by the speaker and is optimal when the speaker develops an unambiguous message protocol with respect to $F$ given $Y$, $\Ladapt$ is null when the listener is optimal, otherwise it is $>0$, ie. sub-optimal. Therefore, we recover the key ingredients of the Loss Decomposition: when the listener is optimal, speaker's loss is limited to $\Linfo$, when the listener is not optimal, the speaker has the additional task to help the listener matching its optimum.







\subsubsection{Case of the Discrimination Game}
\label{app:sec:discri}

Recall that in a discrimination game, as in a reconstruction game, the speaker observes an input, $x$ and sends a message $m$ to the listener. The listener is then provided with both the message $m$, and a list of $N + 1$ candidate inputs, containing input $x$, along with $N$ other inputs, or \textit{distractors}. The goal of the listener is then to give the index of the candidate that corresponds to the actual input.

To formally define discrimination games as instances of the general Lewis game described above, we define $X_1, \ldots, X_N$ to be i.i.d. samples from the inputs distribution $p$. These inputs will be used as the distractors. We additionally set $X_0 = X$. We then define a random permutation $\Sigma$, drawn uniformly from the set of $N + 1$ element permutations, and independently from all other random variables. We then set our auxiliary input $Y = (X_{\Sigma(0)}, \ldots, X_{\Sigma(N)})$, which provides the listener with a permuted list, containing both the correct input at a random position, as well as the distractors. Finally, we set the feature to be predicted as $F = \Sigma^{-1}(0)$. The task of the listener becomes to identify the index of the correct input among all distractors, and we recover a discrimination game.

\subsection{Speaker and Listener Optimizing Different Rewards}
\label{app:sec:asym_reward}

In this paper, we only discuss the case where the agents are fully cooperative, ie. they are optimizing exactly the same reward. When the agents are not aligned on the same objective, the system should be decoupled and an additional \textit{alignement bias} is added to the loss of the speaker. For example, in the reconstruction game where the speaker is optimizing a general reward $r_{\phi}(x,m)=-D(\mathbf{1}_{x},\listener(\cdot|m))+K$ and the listener a cross-entropy loss, the system becomes:

\begin{align}
\left\{
    \begin{array}{ll}
        \mathcal{L}_{\theta} &= \mathbb{E}_{x \sim p, m \sim \speaker(\cdot|x)}[D(\mathbf{1}_{x},\listener(\cdot|m))] - K \\
        \mathcal{L}_{\phi} &= - \mathbb{E}_{x \sim p, m \sim \speaker(\cdot|x)}[\log\listener(x|m)]\cdot
    \end{array}
\right.
\label{eq:system_2}
\end{align}

where $\mathcal{L}_{\theta}$ is the speaker's loss and $\mathcal{L}_{\phi}$ the listener's loss.

By denoting $\rho_{\phi^{*}}(\cdot|m)$ the optimal listener for all $m$ with respect to $\mathcal{L}_{\theta}$ (which is fully determined by the speaker's posterior and $D$) and $\listeneropt(\cdot|m)$ the optimal listener for all $m$ with respect to $\mathcal{L}_{\phi}$ (in this case, the speaker posterior), the speaker loss now decomposes into:

\begin{align*}
    \mathcal{L}_{\theta}&=\underbrace{\mathbb{E}_{m \sim \speaker, x \sim \listeneropt(.|m)}[D(\mathbf{1}_{x}||\listeneropt(\cdot|m))]}_{\Linfo} + \underbrace{\mathbb{E}_{m \sim \speaker, x \sim \listeneropt(.|m)}[D(\mathbf{1}_{x}||\rho_{\phi}(.|m)) - D(\mathbf{1}_{x}||\rho_{\phi^{*}}(\cdot|m))]}_{\Ladapt} \\
    &+ \underbrace{\mathbb{E}_{m \sim \speaker, x \sim \listeneropt(.|m)}[D(\mathbf{1}_{x}||\rho_{\phi^{*}}(.|m)) - D(\mathbf{1}_{x}||\listeneropt(\cdot|m))]}_{\textrm{alignment bias}} - K
\end{align*}

Compared to the standard decomposition, there is an additional term, that we name the \textit{alignment bias}, linked to the gap between the listener optimum of $\mathcal{L_{\theta}}$ and the listener optimum of $\mathcal{L_{\phi}}$. If those optima are close, the amplitude of this term is negligible compared to $\Linfo$ and $\Ladapt$. If those optima are very different (eg. competitive game), the information and co-adaptation terms could have a significantly smaller amplitude compared to the \textit{alignment bias}. We leave to future work the theoretical study of this \textit{alignment bias} which echoes some empirical studies~\citep{noukhovitch2021emergent}.

\section{Method: Additional Computations}
\label{app:method_comp}

In Section~\ref{sec:task_balance_method}, we propose a protocol to balance the importance of the information and co-adaptation losses in the speaker's training loss. To do so, we use the probe listener's estimate of the speaker's posterior on the train set $\probe^{\mathrm{train}}(x|m)=\log\probe^{\mathrm{train}}(x|m)$ and build the following reward:

\begin{equation*}
      r_{\phi}(x,m;\alpha) = (1-2\alpha) \times \underbrace{\log\probe^{\mathrm{train}}(x|m)}_{\text{probe listener reward}}  \; + \; \alpha \times \underbrace{ \log\listener(x|m)}_{\text{standard listener reward}}
\end{equation*}
where $\alpha$ is a weight in $[0;0.5]$.

The loss equality defined in Section~\ref{sec:task_balance_method} is then recovered with the following computations:

\begin{align*}
    \mathcal{L}_{\theta}(\alpha) &= -\mathbb{E}_{x\sim p, m \sim \pi_{\theta}(.|x)}[r_{\phi}(x,m;\alpha)] \\
     &= -\mathbb{E}_{x\sim p, m \sim \pi_{\theta}(.|x)}[(1-2\alpha) \times \log\probe^{\mathrm{train}}(x|m)  \; + \; \alpha \times \log\listener(x|m)] \\ 
     &= -(1-2\alpha)\mathbb{E}_{x\sim p, m \sim \pi_{\theta}(.|x)}[\log\probe^{\mathrm{train}}(x|m)] - \alpha\mathbb{E}_{x\sim p, m \sim \pi_{\theta}(.|x)}[\log\listener(x|m)] \\
     &= (1-2\alpha)\Linfoestimate^{\mathrm{train}} + \alpha(\Linfoestimate^{\mathrm{train}}+\Ladaptestimate^{\mathrm{train}}) \\ 
     \mathcal{L}_{\theta}(\alpha) &= (1-\alpha) \Linfoestimate^{\mathrm{train}} + \alpha \Ladaptestimate^{\mathrm{train}}
\end{align*}

\paragraph{Remark} In the paper, we only consider the case $\alpha \in [0;0.5]$ and do not explore larger values of $\alpha$. Indeed, controlling the co-adaptation rate $\alpha$ is made by re-weighting $\Linfoestimate^{\mathrm{train}}$ (estimated with a probe listener). However, two issues occur when $\alpha>0.5$:
\begin{itemize}
    \item First, the goal of computing $\Linfoestimate^{\mathrm{train}}$ is to indirectly balance the weight of the training information loss $\Linfo^{train}$.
    By taking the loss of the probe listener close to optimality, we get an upper bound estimate $\Linfoestimate^{\mathrm{train}}$ of the training information loss $\Linfo^{\mathrm{train}}$. Therefore, it theoretically ensures that we minimize $\Linfo^{\mathrm{train}}$ when optimizing $\Linfoestimate^{\mathrm{train}}$. However, when $\alpha>0.5$, the weight of $\Linfoestimate^{\mathrm{train}}$ is negative. In this case, since $\Linfoestimate^{\mathrm{train}}$ is an upper bound of $\Linfo^{\mathrm{train}}$, we do not have the guarantee that the speaker minimizes $-\Linfo^{\mathrm{train}}$ anymore.
    \item Second, we empirically experimented $\alpha>0.5$ even if theoretical conditions are not reached. In practice, if the system converged for values of $\alpha$ closed to $0.5$, the system quickly became unstable for larger values of $\alpha$. Our main hypothesis is that the speaker cannot start structuring its messages when the weight of $\Ladaptestimate^{\mathrm{train}}$ is too strong. Indeed, agents start with random weights. It implies that, at the beginning of the training, if the weight of $\Ladaptestimate^{\mathrm{train}}$ is too strong, it pressures the speaker to have an almost uniform posterior, ie. to develop a fully ambiguous language. In short, if $\alpha$ is too large, the speaker has too little pressure on developing meaningful messages and therefore succeeding in the communication task.
\end{itemize}

\section{Regularization}

We here provide:
\begin{itemize}
    \item the parameters used for the listener's regularization (Appendix~\ref{app:reg_params})
    \item the results obtained when regularizing the speaker (Appendix~\ref{app:reg_speaker})
\end{itemize}

\subsection{Parameters of the Listener's Regularization}
\label{app:reg_params}

Regularization parameters have been tuned in order to get the best average generalization scores while having a convergence success rate greater or equal to $75\%$. When regularizing with the layer normalization (noted \textit{No LN.} in Table~\ref{tab:reg_listener}), we remove the layer normalization applied of the listener's LSTM cell. Dropout rate is set to $0.2$ and weight decay penalty is set to $0.01$ both when layer normalization is kept (noted \textit{Weight decay} in Table~\ref{tab:reg_listener}) and when layer normalization is removed (noted \textit{No LN. + WD} in Table~\ref{tab:reg_listener}).

\subsection{Comparison with Speaker's Regularization}
\label{app:reg_speaker}

\paragraph{Parameters} For the sake of completeness, we also study the impact of regularizing the speaker. Here, we only report the results with the weight decay penalty. Indeed, removing the layer normalization makes the training slow and unstable while results with dropout are worse than those with weight decay. Weight decay penalty has been fine-tuned to $0.005$ to get the best average generalization performances while having $>75\%$ successful experiments.

\paragraph{Results} In Table~\ref{tab:compa_res_reg_sp}, we compare the generalization and compositionality of emergent languages with and without regularization applied on the speaker. First, when we regularize the speaker without any regularization on the listener, we see that the gain of generalization and compositionality is negligible and inferior to the gain obtained when regularizing the listener. Moreover, we note that when we regularize both the speaker and the listener, scores of generalization and compositionality are similar to those obtained when only regularizing the listener. It suggests that regularizing the speaker has little impact on generalization and compositionality.

These results support the claim of Section~\ref{sec:compa_reg}: the listener is the main contributor of the co-adaptation overfitting in the reconstruction game.

\begin{table}[h!]
    \centering
    \hfill
    \begin{tabular}{lcc}
    \toprule
     \textbf{No Speaker reg.} &  Gen. $\uparrow$ & Compo. $\uparrow$  \\ 
     \midrule \midrule
         Continuous & $0.58_{\pm0.05}$ & $0.22_{\pm0.02}$  \\
         \midrule
         No LN. & $0.70_{\pm0.03}$ & $0.24_{\pm0.02}$ \\
         Weight decay &  $0.72_{\pm0.03}$ & $0.25_{\pm0.03}$ \\
         No LN. + WD & $0.87_{\pm0.07}$ & $0.30_{\pm0.03}$ \\ 
     \bottomrule
    \end{tabular}
    \hfill
    \begin{tabular}{lcc}
    \toprule
     \textbf{Speaker with WD} &  Gen. $\uparrow$ & Compo. $\uparrow$  \\ 
     \midrule \midrule
         Continuous & $0.62_{\pm0.02}$ & $0.22_{\pm 0}.03$ \\
         \midrule
         No LN. & $0.68_{\pm 0.07}$ & $0.23_{\pm 0.01}$  \\
         Weight decay &  $0.74_{\pm 0.05}$ & $0.26_{\pm 0.04}$  \\
         No LN. + WD & $0.82_{\pm 0.07}$ & $0.32_{\pm 0.04}$  \\
     \bottomrule
    \end{tabular}
    \hfill
    \caption{Performance comparison: (left) without speaker regularization ; (right) with speaker regularization. Weight decay penalty on the speaker is set to $0.005$. Parameters of regularization methods for the listener are reported in Appendix~\ref{app:reg_params}.}
    \label{tab:compa_res_reg_sp}
\end{table}

\section{Image Discrimination Games}
\label{app:imag}

We here complete Section~\ref{sec:scaling_to_image} by presenting the rules and experimental settings of the image discrimination game (Appendix~\ref{app:image_expe}), reporting the results of compositionality (Appendix~\ref{app:top_sim_images}) and completing generalization results of Table~\ref{tab:reg_listener} with regularization experiments (Appendix~\ref{app:top_gen_images}).

\subsection{Experimental Settings}
\label{app:image_expe}

For the implementation of the image discrimination game, we mostly follow the protocol proposed by \citep{chaabouni2022emergent}.

\subsubsection{Game Rules and notations}

In the Lewis image discrimination game, the speaker observes an image. Then, the speaker sends a descriptive message to the listener. Based on this message, the listener should retrieve the correct image among a set of candidates. 

Formally, the image observed by the speaker is denoted by $x$ and belongs to a set $\mathcal{X}$. The intermediate message sent by the speaker is denoted by $m$ and belongs to a set a potential messages $\mathcal{M}$. The speaker follows a policy $\speaker$ which samples a message $m$ with probability $\speaker(m|x)$ conditioned on image $x$. The listener encodes the message $m$ into a representation $t_{\phi}(m)$. The set of candidates received by the listener are denoted $\mathcal{C}$ and the listener encodes each candidates $x' \in \mathcal{C}$ by a representation $t_{\phi}(x')$. The probability of a candidate $x'$ to be the correct image is : $\listener(x'|m,\mathcal{C})$. It is obtained by comparing the message encoding $t_{\phi}(m)$ with the image encoding $t_{\phi}(x')$ of all candidates.

\subsubsection{Environment}

\paragraph{Datasets} We perform the discrimination game on ImageNet~\citep{deng2009imagenet,russakovsky2015imagenet} and CelebA~\citep{liu2015deep}. We work with image pre-processed encodings $f(x)$ of size $2048$ that have been open-sourced by \citep{chaabouni2022emergent}. In the two datasets, each image has been center-cropped and processed by a ResNet-50 encoder pretrained on ImageNet with the self-supervised method BYOL~\citep{grill2020bootstrap}.

\paragraph{Train/val/test splits} For building our custom training sets, we first considered the splits provided by \citep{chaabouni2022emergent}. From the respective $1400k$ and $200k$ labelled images of ImageNet and CelebA, they slitted the dataset in train, validation and test with the ratio $80/10/10$.

To test agents generalization capacities, we also build subsets of the training set provided by \citep{chaabouni2022emergent}: ImageNet $\frac{1}{20}$, ImageNet $\frac{1}{100}$, CelebA $\frac{1}{20}$ and CelebA $\frac{1}{100}$. For each of those sub-training sets, we randomly selected a small fraction of the training set, approximatively corresponding to $1/20$-th and $1/100$-th of the total training set. The corresponding number of samples are reported in Table~\ref{tab:training_samples}.

\begin{table}[h!]
\centering
        \begin{tabular}{lcc}
        \toprule
          \multicolumn{1}{l}{} & \multicolumn{2}{c}{Training samples} \\ 
          \midrule
          \midrule
          \textbf{CelebA}      & $1/20$ & $1/100$   \\
          \midrule
              & $8492$        & $2123$  \\ 
              \bottomrule
        \end{tabular}
        \begin{tabular}{lcc}
        \toprule
          \multicolumn{1}{l}{} & \multicolumn{2}{c}{Training samples} \\ 
          \midrule
          \midrule
          \textbf{ImageNet}      & $1/20$ & $1/100$   \\
          \midrule
           & $50732$ & $12683$  \\ 
        \bottomrule
        \end{tabular}
        \caption{Number of training samples for the four training subsets considered: ImageNet $\frac{1}{20}$, ImageNet $\frac{1}{100}$, CelebA $\frac{1}{20}$ and CelebA $\frac{1}{100}$}
\label{tab:training_samples}
\end{table}

All our experiments on images are run with those $4$ small training sets. We keep the original validation and test sets from \citep{chaabouni2022emergent}.

\subsubsection{Agent Models}

\paragraph{Speaker model} The speaker is a neural network that takes the pre-processed representation of an image $f(x)$ as input of size $2048$ and returns a message $m=(m_{i})_{1 \leq i \leq T}$ of length $T$. 

The speaker follows a recurrent policy: given the image representation $f(x)$, it samples for all $t \in [1,T]$ a token $m_{t}$ with probability $\speaker(m_{t}|m_{<t},f(x))$.
The image representation $f(x)$ is first projected by a linear layer to get an object embedding of size $256$ that is used to initialize a LSTM of size $256$ with layer normalization. At each time step, the LSTM's output is fed into a linear layer of size $|\mathcal{V}|$, followed by a softmax, to produce $\speaker(m_{t}|m_{<t},f(x))$. 

In our experiments, the following parameters have been chosen: $T=10,|V|=10$ meaning that the message space is of size $10^{10}$ preventing any channel capacity bottleneck.

\paragraph{Listener model} The listener is a neural network that takes the speaker's message $m$ and a set of image candidates $\mathcal{C}$ containing the target image $x$ and outputs the probability for each candidate $x' \in \mathcal{C}$ to be the target image $x$.

The listener is composed of two modules: one that encodes the message ; the other that encodes images. For a message $m = (m_{1},...,m_{T})$, the listener passes each symbol $m_{t}$ through an embedding layer of dimension $256$ followed by a LSTM of size $256$ with layer normalization. The final recurrent state $h^{1}_{T}$ is then passed to a linear layer that produces the image encoding $t_{\phi}(m)$ of size $256$. In parallel, each candidate $x'$ is first pre-processed by $f$ and then passed through a linear layer producing an image encoding $t_{\phi}(x')$ of size $256$.

The message representation $t_{\phi}(m)$ is then compared to each candidate representation $t_{\phi}(x')$ with the following score function: $\mathrm{score}(m,x',\phi):=t_{\phi}(m)t_{\phi}(x')^{T}$. Note that contrary to \citep{chaabouni2022emergent}, we rather use a dot-product score function~\citep{lazaridou2018emergence} instead of a cosine similarity because we empirically got better results and more stable trainings. The probability distribution over the candidates $\mathcal{C}$ of being the target image $x$ is then obtained by normalizing the scores with a softmax. This probability distribution is denoted by $\listener(\cdot | m, \mathcal{C})$ and the listener guess is $\hat{x}=\underset{x'}{\mathrm{argmax}}\listener(x' | m, \mathcal{C})$.

In our experiments, the number of candidates is $|\mathcal{C}|=1000$.

\subsubsection{Agents Training}

We follow the same principle as in the reconstruction game: the listener is trained to best predict the target image among the set of candidates, while the speaker takes the opposite of the listener's loss as reward:

\paragraph{Listener loss} The listener is trained to predict the target image among the set of candidates $\mathcal{C}$. When receiving a batch of inputs $x$, a set of candidates $\mathcal{C}$ is sampled for each input $x$. The sampling is uniform without replacement over $\mathcal{X}-\{x\}$ meaning that the target image $x$ cannot be duplicated into the candidates. The listener is then trained to optimized the average InfoNCE loss~\citep{oord2018representation}:
\begin{align*}
    \mathcal{L}_{\phi}=\sum_{x \in \mathrm{batch}}-\log \listener(x | m, \mathcal{C})
\end{align*}

\paragraph{Speaker loss} When the speaker observes an image $x$, sends a message $m$ and the listener has to choose among a set of candidates $\mathcal{C}$, the speaker's reward is defined as: 
\begin{align*}
    r_{\phi,\mathcal{C}}(x,m)=\log \listener(x | m, \mathcal{C})
\end{align*}

The speaker is trained to maximize its cumulative reward: $\mathbb{E}_{x,m,\mathcal{C}}[\log \listener(x | m, \mathcal{C})]$ which means that the speaker and the listener have the same loss.

\paragraph{Optimization} The agents are optimized using \texttt{Adam}~\citep{adam-opt} with $\beta_1=0.9$ and $\beta_2=0.999$. The speaker's learning rate is $5\cdot 10^{-4}$ while the listener's learning rate is $1\cdot 10^{-3}$.
Agents are trained on batches of size of $2048$. For the speaker, we use policy gradient~\citep{sutton2000policy}, with a baseline computed as the average reward within the minibatch, and we add an entropy regularization of $0.01$ to the speaker's loss~\citep{williams1991function}. 




\subsection{Topographic Similarity Results}
\label{app:top_sim_images}

We report results of topographic similarity for experiments of Section~\ref{sec:scaling_to_image}. To be complete, we add the scores when applying listener regularization (corresponding generalization performances are reported in Appendix~\ref{app:top_gen_images}).

Scores of topographic similarity are reported in Table~\ref{tab:top_sim_image}. Here, the distance used to compare images is the cosine distance between the vector representations of the ResNet-50 encoder pretrained on ImageNet. The distance used to compare messages remains the edit-distance. As mentioned in the main paper, we can see that there is not any compositionality trend when agents communicate about images. Moreover, when comparing with Table~\ref{tab:more_res_image} that reports generalization performances, we see that gains of generalization do not correlate with gains of topographic similarity. It suggests that the topographic similarity does not capture agents' language structure in image based settings, as already observed in previous work~\citep{chaabouni2022emergent,andreas2019measuring}. 

\begin{table}[h!]
      \centering
        \begin{tabular}{lcc}
        \toprule
          \multicolumn{1}{l}{} & \multicolumn{2}{c}{Topographic similarity $\uparrow$} \\ 
          \midrule
          \midrule
          \textbf{CelebA}      & $1/20$ & $1/100$   \\
          \midrule
          Continuous    & $\mathbf{0.28_{\pm0.03}}$ & $\mathbf{0.32_{\pm0.03}}$  \\ 
          No LN.    & $0.26_{\pm0.04}$          & $0.29_{\pm0.03}$  \\ 
          No LN. + WD & --          & $0.30_{\pm0.03}$  \\ 
          Weight decay    & $0.27_{\pm0.03}$          & $0.28_{\pm0.04}$  \\
          Early stopping    & $0.27_{\pm0.04}$ & $0.30_{\pm0.03}$  \\
        \bottomrule
        \end{tabular}
        \begin{tabular}{lcc}
        \toprule
          \multicolumn{1}{l}{} & \multicolumn{2}{c}{Topographic similarity $\uparrow$} \\ 
          \midrule
          \midrule
          \textbf{ImageNet}      & $1/20$ & $1/100$   \\
          \midrule
          Continuous    & $0.17_{\pm0.03}$& $0.17_{\pm0.03}$  \\ 
          No LN.    & $0.18_{\pm0.01}$          & $0.16_{\pm0.02}$  \\ 
          No LN. + WD    & $\mathbf{0.19_{\pm0.03}}$          & $\mathbf{0.21_{\pm0.03}}$  \\ 
          Weight decay & $0.17_{\pm0.02}$          & $0.15_{\pm0.02}$  \\ 
          Early stopping    & $0.18_{\pm0.04}$ & $0.20_{\pm0.03}$ \\
        \bottomrule
        \end{tabular}
        \caption{Topographic similarity of emergent languages in the image discrimination game where images are compared with a cosine similarity. \textit{No LN.} refers to the removal of the layernorm on the listener's LSTM cell ; \textit{Weight decay} to the addition of weight decay on the listener with penalty equal to $0.01$ ; \textit{No LN. + WD} refers to the removal of layernorm and addition of weight decay on the listener. No result for \textit{No LN. + WD} are reported with Celeba $\frac{1}{20}$ because experiments did not converge with the regularization parameters chosen.}
        \label{tab:top_sim_image}
\end{table}

In addition, we also test whether scores of topographic similarities are improved when using another distance to compare images. In Table~\ref{tab:top_sim_image_att}, we use the attributes provided in CelebA to compare the images. The distance between two images is computed as $1-\text{(propotion of common attributes)}$. For the message comparison, we keep the edit-distance. Once again, no topographic similarity trends emerge, sustaining results already observed in \citep{chaabouni2022emergent}. 

\begin{table}[h!]
      \centering
        \begin{tabular}{lcc}
        \toprule
          \multicolumn{1}{l}{} & \multicolumn{2}{c}{Topographic similarity (with attributes) $\uparrow$} \\ 
          \midrule
          \midrule
          \textbf{CelebA}      & $1/20$ & $1/100$   \\
          \midrule
          Continuous    & $0.13_{\pm0.02}$ & $\mathbf{0.15_{\pm0.03}}$  \\ 
          No LN.    & $\mathbf{0.14_{\pm0.02}}$          & $0.14_{\pm0.03}$  \\ 
          No LN. + WD & --          & $\mathbf{0.15_{\pm0.04}}$  \\ 
          Weight decay    & $\mathbf{0.14_{\pm0.02}}$          & $\mathbf{0.15_{\pm0.01}}$  \\
          Early stopping    & $0.13_{\pm0.02}$ & $\mathbf{0.15_{\pm0.02}}$  \\
        \bottomrule
        \end{tabular}
        \caption{Topographic similarity of emergent languages in the image discrimination game where images are compared with CelebA attributes. \textit{No LN.} refers to the removal of the layer normalization on the listener's LSTM cell ; \textit{Weight decay} to the addition of weight decay on the listener with penalty equal to $0.01$ ; \textit{No LN. + WD} refers to the removal of layer normalization and addition of weight decay on the listener. No result for \textit{No LN. + WD} are reported with Celeba $\frac{1}{20}$ because experiments did not converge with the regularization parameters chosen.}
        \label{tab:top_sim_image_att}
\end{table}

\subsection{More Results with Listener Regularization}
\label{app:top_gen_images}

To complete the generalization scores of Table~\ref{tab:reg_listener} in the main paper, we report in Table~\ref{tab:more_res_image} the generalization scores in the image discrimination game for various regularization methods applied on the listener. We observe the same trends as in the reconstruction game. Indeed, listener regularization consistently improves the performances. It means, that a large gain of performance can be obtained in those games by regularizing the listener. The \textit{Early stopping listener} remains a top line in image based experiments.

\begin{table}[h!]
      \centering
        \begin{tabular}{lcc}
        \toprule
          \multicolumn{1}{l}{} & \multicolumn{2}{c}{Generalization $\uparrow$} \\ 
          \midrule
          \midrule
          \textbf{CelebA}      & $1/20$ & $1/100$   \\
          \midrule
          Continuous    &  $0.67_{\pm0.02}$          & $0.39_{\pm0.07}$  \\ 
          No LN.    & $0.67_{\pm0.03}$           & $0.44_{\pm0.02}$  \\ 
          No LN. + WD  & --          & $0.50_{\pm0.07}$  \\
          Weight decay    & $0.77_{\pm0.04}$        &  $0.60_{\pm0.06}$ \\
          Early stopping    & $\mathbf{0.80_{\pm0.03}}$ & $\mathbf{0.69_{\pm0.04}}$  \\
        \bottomrule
        \end{tabular}
        \begin{tabular}{lcc}
        \toprule
          \multicolumn{1}{l}{} & \multicolumn{2}{c}{Generalization $\uparrow$} \\ 
          \midrule
          \midrule
          \textbf{ImageNet}     & $1/20$ & $1/100$   \\
          \midrule
          Continuous   & $0.77_{\pm0.01   }$ & $0.51_{\pm0.03}$ \\ 
          No LN.   & $0.77_{\pm0.01}$          & $0.53_{\pm0.03}$   \\ 
          No LN. + WD  & $0.75_{\pm0.01}$          &
          $0.59_{\pm0.04}$  \\ 
          Weight decay    & $0.79_{\pm0.03}$ & $0.62_{\pm0.02}$  \\ 
          Early stopping    & $\mathbf{0.81_{\pm0.01}}$ & $\mathbf{0.64_{\pm0.01}}$ \\
        \bottomrule
        \end{tabular}
        \caption{Comparison of generalization performances between the \textit{Continuous listener}, \textit{Early stopping listener} and listeners with regularization on the image discrimination game. \textit{No LN.} refers to the removal of the layer normalization on the listener's LSTM cell ; \textit{Weight decay} to the addition of weight decay on the listener with penalty equal to $0.01$ ; \textit{No LN. + WD} refers to the removal of layer normalization and addition of weight decay on the listener. No result for \textit{No LN. + WD} are reported with Celeba $\frac{1}{20}$ because experiments did not converge with the regularization parameters chosen.}
        \label{tab:more_res_image}
\end{table}

\clearpage



%% file: main.bbl
\begin{thebibliography}{85}
\providecommand{\natexlab}[1]{#1}
\providecommand{\url}[1]{\texttt{#1}}
\expandafter\ifx\csname urlstyle\endcsname\relax
  \providecommand{\doi}[1]{doi: #1}\else
  \providecommand{\doi}{doi: \begingroup \urlstyle{rm}\Url}\fi

\bibitem[Andreas(2019)]{andreas2019measuring}
Jacob Andreas.
\newblock Measuring compositionality in representation learning.
\newblock In \emph{Proc. of International Conference on Learning
  Representations (ICLR)}, 2019.

\bibitem[Ba et~al.(2016)Ba, Kiros, and Hinton]{ba2016layer}
Jimmy~Lei Ba, Jamie~Ryan Kiros, and Geoffrey~E. Hinton.
\newblock Layer normalization.
\newblock \emph{arXiv preprint arXiv:1607.06450}, 2016.

\bibitem[Baroni(2020)]{baroni2020}
Marco Baroni.
\newblock Linguistic generalization and compositionality in modern artificial
  neural networks.
\newblock \emph{Philosophical Transactions of the Royal Society B: Biological
  Sciences}, 375\penalty0 (1791):\penalty0 20190307, 2020.
\newblock \doi{10.1098/rstb.2019.0307}.

\bibitem[Barsalou(2008)]{barsalou2008grounded}
Lawrence~W Barsalou.
\newblock Grounded cognition.
\newblock \emph{Annu. Rev. Psychol.}, 59:\penalty0 617--645, 2008.

\bibitem[Bickerton(2007)]{bickerton2007language}
Derek Bickerton.
\newblock Language evolution: A brief guide for linguists.
\newblock \emph{Lingua}, 117\penalty0 (3):\penalty0 510--526, 2007.

\bibitem[Bickerton(2014)]{bickerton2014more}
Derek Bickerton.
\newblock More than nature needs.
\newblock In \emph{More than Nature Needs}. Harvard University Press, 2014.

\bibitem[Brennan and Clark(1996)]{brennan1996conceptual}
Susan~E Brennan and Herbert~H Clark.
\newblock Conceptual pacts and lexical choice in conversation.
\newblock \emph{Journal of experimental psychology: Learning, memory, and
  cognition}, 22\penalty0 (6):\penalty0 1482, 1996.

\bibitem[Brighton and Kirby(2006)]{brighton2006}
Henry Brighton and Simon Kirby.
\newblock {Understanding Linguistic Evolution by Visualizing the Emergence of
  Topographic Mappings}.
\newblock \emph{Artificial Life}, 12\penalty0 (2):\penalty0 229--242, 04 2006.
\newblock ISSN 1064-5462.
\newblock \doi{10.1162/artl.2006.12.2.229}.

\bibitem[Chaabouni et~al.(2019)Chaabouni, Kharitonov, Dupoux, and
  Baroni]{chaabouni2019anti}
Rahma Chaabouni, Eugene Kharitonov, Emmanuel Dupoux, and Marco Baroni.
\newblock Anti-efficient encoding in emergent communication.
\newblock In \emph{Proc. of Advances in Neural Information Processing Systems
  (NeurIPS)}, 2019.

\bibitem[Chaabouni et~al.(2020)Chaabouni, Kharitonov, Bouchacourt, Dupoux, and
  Baroni]{chaabouni2020compositionality}
Rahma Chaabouni, Eugene Kharitonov, Diane Bouchacourt, Emmanuel Dupoux, and
  Marco Baroni.
\newblock Compositionality and generalization in emergent languages.
\newblock In \emph{Proc. of the Association for Computational Linguistics
  (ACL)}, 2020.

\bibitem[Chaabouni et~al.(2021)Chaabouni, Kharitonov, Dupoux, and
  Baroni]{chaabouni2021communicating}
Rahma Chaabouni, Eugene Kharitonov, Emmanuel Dupoux, and Marco Baroni.
\newblock Communicating artificial neural networks develop efficient
  color-naming systems.
\newblock \emph{Proceedings of the National Academy of Sciences}, 118\penalty0
  (12), 2021.

\bibitem[Chaabouni et~al.(2022)Chaabouni, Strub, Altch{\'e}, Tarassov, Tallec,
  Davoodi, Mathewson, Tieleman, Lazaridou, and Piot]{chaabouni2022emergent}
Rahma Chaabouni, Florian Strub, Florent Altch{\'e}, Eugene Tarassov, Corentin
  Tallec, Elnaz Davoodi, Kory~Wallace Mathewson, Olivier Tieleman, Angeliki
  Lazaridou, and Bilal Piot.
\newblock Emergent communication at scale.
\newblock In \emph{Proc. of International Conference on Learning
  Representations (ICLR)}, 2022.

\bibitem[Christiansen and Kirby(2003)]{christiansen2003language}
Morten~H Christiansen and Simon Kirby.
\newblock Language evolution: Consensus and controversies.
\newblock \emph{Trends in cognitive sciences}, 7\penalty0 (7):\penalty0
  300--307, 2003.

\bibitem[Clyne(1992)]{clyne1992linguistic}
Michael Clyne.
\newblock Linguistic and sociolinguistic aspects of language contact,
  maintenance and loss.
\newblock \emph{Maintenance and loss of minority languages}, 1:\penalty0 17,
  1992.

\bibitem[Cogswell et~al.(2019)Cogswell, Lu, Lee, Parikh, and
  Batra]{cogswell2020emergence}
Michael Cogswell, Jiasen Lu, Stefan Lee, Devi Parikh, and Dhruv Batra.
\newblock Emergence of compositional language with deep generational
  transmission.
\newblock \emph{arXiv preprint arXiv:1904.09067}, 2019.

\bibitem[Cornish et~al.(2017)Cornish, Dale, Kirby, and
  Christiansen]{cornish2017sequence}
Hannah Cornish, Rick Dale, Simon Kirby, and Morten~H Christiansen.
\newblock Sequence memory constraints give rise to language-like structure
  through iterated learning.
\newblock \emph{PloS one}, 12\penalty0 (1):\penalty0 e0168532, 2017.

\bibitem[Crawford and Sobel(1982)]{crawford1982strategic}
Vincent~P Crawford and Joel Sobel.
\newblock Strategic information transmission.
\newblock \emph{Econometrica: Journal of the Econometric Society}, pages
  1431--1451, 1982.

\bibitem[Cultural Intelligence~Team et~al.(2022)Cultural Intelligence~Team,
  Bhoopchand, Brownfield, Collister, Dal~Lago, Edwards, Everett, Frechette,
  Gitahy~Oliveira, Hughes, et~al.]{cultural2022learning}
General Cultural Intelligence~Team, Avishkar Bhoopchand, Bethanie Brownfield,
  Adrian Collister, Agustin Dal~Lago, Ashley Edwards, Richard Everett,
  Alexandre Frechette, Yanko Gitahy~Oliveira, Edward Hughes, et~al.
\newblock Learning robust real-time cultural transmission without human data.
\newblock \emph{arXiv preprint arXiv:2203.00715}, 2022.

\bibitem[Deng et~al.(2009)Deng, Dong, Socher, Li, Li, and
  Fei-Fei]{deng2009imagenet}
Jia Deng, Wei Dong, Richard Socher, Li-Jia Li, Kai Li, and Li~Fei-Fei.
\newblock Imagenet: A large-scale hierarchical image database.
\newblock In \emph{Proc. of Conference on Computer Vision and Pattern
  Recognition (CVPR)}, 2009.

\bibitem[Dess{\`\i} and Baroni(2019)]{dessi2019cnns}
Roberto Dess{\`\i} and Marco Baroni.
\newblock Cnns found to jump around more skillfully than rnns: Compositional
  generalization in seq2seq convolutional networks.
\newblock \emph{arXiv preprint arXiv:1905.08527}, 2019.

\bibitem[Dess{\`\i} et~al.(2021)Dess{\`\i}, Kharitonov, and
  Marco]{dessi2021interpretable}
Roberto Dess{\`\i}, Eugene Kharitonov, and Baroni Marco.
\newblock Interpretable agent communication from scratch (with a generic visual
  processor emerging on the side).
\newblock \emph{Advances in Neural Information Processing Systems}, 34, 2021.

\bibitem[Ellis(2008)]{ellis2008dynamics}
Nick~C Ellis.
\newblock The dynamics of second language emergence: Cycles of language use,
  language change, and language acquisition.
\newblock \emph{The modern language journal}, 92\penalty0 (2):\penalty0
  232--249, 2008.

\bibitem[Evtimova et~al.(2018)Evtimova, Drozdov, Kiela, and
  Cho]{evtimova2018emergent}
Katrina Evtimova, Andrew Drozdov, Douwe Kiela, and Kyunghyun Cho.
\newblock Emergent communication in a multi-modal, multi-step referential game.
\newblock In \emph{Proc. of International Conference on Learning
  Representations (ICLR)}, 2018.

\bibitem[Fay and Ellison(2013)]{fay2013cultural}
Nicolas Fay and T~Mark Ellison.
\newblock The cultural evolution of human communication systems in different
  sized populations: usability trumps learnability.
\newblock \emph{PloS one}, 8\penalty0 (8):\penalty0 e71781, 2013.

\bibitem[Galantucci and Garrod(2011)]{galantucci2011experimental}
Bruno Galantucci and Simon Garrod.
\newblock Experimental semiotics: a review.
\newblock \emph{Frontiers in human neuroscience}, 5:\penalty0 11, 2011.

\bibitem[Galke et~al.(2022)Galke, Ram, and Raviv]{galke2022emergent}
Lukas Galke, Yoav Ram, and Limor Raviv.
\newblock Emergent communication for understanding human language evolution:
  What's missing?
\newblock In \emph{Emergent Communication Workshop at ICLR}, 2022.

\bibitem[Gary~Lupyan(2010)]{Lupyan2010}
Rick~Dale Gary~Lupyan.
\newblock Language structure is partly determined by social structure.
\newblock \emph{PLoS ONE 5}, 1, 2010.

\bibitem[Graesser et~al.(2019)Graesser, Cho, and
  Kiela]{harding-graesser-etal-2019-emergent}
Laura Graesser, Kyunghyun Cho, and Douwe Kiela.
\newblock Emergent linguistic phenomena in multi-agent communication games.
\newblock In \emph{Proc. of Empirical Methods in Natural Language Processing
  (EMNLP)}, 2019.

\bibitem[Grill et~al.(2020)Grill, Strub, Altch{\'e}, Tallec, Richemond,
  Buchatskaya, Doersch, Avila~Pires, Guo, Gheshlaghi~Azar,
  et~al.]{grill2020bootstrap}
Jean-Bastien Grill, Florian Strub, Florent Altch{\'e}, Corentin Tallec, Pierre
  Richemond, Elena Buchatskaya, Carl Doersch, Bernardo Avila~Pires, Zhaohan
  Guo, Mohammad Gheshlaghi~Azar, et~al.
\newblock Bootstrap your own latent-a new approach to self-supervised learning.
\newblock \emph{Advances in Neural Information Processing Systems},
  33:\penalty0 21271--21284, 2020.

\bibitem[Guo et~al.(2021)Guo, Ren, Mathewson, Kirby, Albrecht, and
  Smith]{guo2021expressivity}
Shangmin Guo, Yi~Ren, Kory Mathewson, Simon Kirby, Stefano~V Albrecht, and
  Kenny Smith.
\newblock Expressivity of emergent language is a trade-off between contextual
  complexity and unpredictability.
\newblock In \emph{Proc. of International Conference on Learning
  Representations (ICLR)}, 2021.

\bibitem[Harnad(1990)]{harnad1990symbol}
Stevan Harnad.
\newblock The symbol grounding problem.
\newblock \emph{Physica D: Nonlinear Phenomena}, 42\penalty0 (1-3):\penalty0
  335--346, 1990.

\bibitem[Harnad et~al.(1976)Harnad, Steklis, and
  Lancaster]{Harnad1976OriginsAE}
Steven~R Harnad, Horst~D Steklis, and Jane~Ed Lancaster.
\newblock Origins and evolution of language and speech.
\newblock \emph{Annals of the New York Academy of Sciences}, 1976.

\bibitem[Havrylov and Titov(2017)]{havrylov2017emergence}
Serhii Havrylov and Ivan Titov.
\newblock Emergence of language with multi-agent games: Learning to communicate
  with sequences of symbols.
\newblock In \emph{Proc. of Advances in Neural Information Processing Systems
  (NeurIPS)}, 2017.

\bibitem[Hinton(1987)]{hinton1987learning}
Geoffrey~E Hinton.
\newblock Learning translation invariant recognition in a massively parallel
  networks.
\newblock In \emph{International Conference on Parallel Architectures and
  Languages Europe}, pages 1--13. Springer, 1987.

\bibitem[Hochreiter and Schmidhuber(1997)]{hochreiter1997long}
Sepp Hochreiter and J{\"u}rgen Schmidhuber.
\newblock Long short-term memory.
\newblock \emph{Neural computation}, 9\penalty0 (8):\penalty0 1735--1780, 1997.

\bibitem[Huttegger et~al.(2014)Huttegger, Skyrms, Tarres, and
  Wagner]{huttegger2014some}
Simon Huttegger, Brian Skyrms, Pierre Tarres, and Elliott Wagner.
\newblock Some dynamics of signaling games.
\newblock \emph{Proceedings of the National Academy of Sciences}, 111\penalty0
  (Supplement 3):\penalty0 10873--10880, 2014.

\bibitem[Kalinowska et~al.(2022)Kalinowska, Davoodi, Strub, Mathewson, Murphey,
  and Pilarski]{kalinowska2022situated}
Aleksandra Kalinowska, Elnaz Davoodi, Florian Strub, Kory Mathewson, Todd
  Murphey, and Patrick Pilarski.
\newblock Situated communication: A solution to over-communication between
  artificial agents.
\newblock In \emph{Emergent Communication Workshop at ICLR 2022}, 2022.

\bibitem[Kharitonov and Baroni(2020)]{kharitonov-baroni-2020-emergent}
Eugene Kharitonov and Marco Baroni.
\newblock Emergent language generalization and acquisition speed are not tied
  to compositionality.
\newblock In \emph{Proc. of the BlackboxNLP Workshop on Analyzing and
  Interpreting Neural Networks for NLP}, 2020.

\bibitem[Kharitonov et~al.(2019)Kharitonov, Chaabouni, Bouchacourt, and
  Baroni]{kharitonov2019egg}
Eugene Kharitonov, Rahma Chaabouni, Diane Bouchacourt, and Marco Baroni.
\newblock Egg: a toolkit for research on emergence of language in games.
\newblock In \emph{Proc. of Empirical Methods in Natural Language Processing
  (EMNLP)}, 2019.

\bibitem[Kharitonov et~al.(2020)Kharitonov, Chaabouni, Bouchacourt, and
  Baroni]{kharitonov2020entropy}
Eugene Kharitonov, Rahma Chaabouni, Diane Bouchacourt, and Marco Baroni.
\newblock Entropy minimization in emergent languages.
\newblock In \emph{Proc. of International Conference on Machine Learning
  (ICML)}, 2020.

\bibitem[Kim and Oh(2021)]{kim2021emergent}
Jooyeon Kim and Alice Oh.
\newblock Emergent communication under varying sizes and connectivities.
\newblock \emph{Proc. of Advances in Neural Information Processing Systems
  (NeurIPS)}, 2021.

\bibitem[Kingma and Ba(2015)]{adam-opt}
Diederik~P Kingma and Jimmy Ba.
\newblock Adam: A method for stochastic optimization.
\newblock In \emph{Proc. of International Conference on Learning
  Representations (ICLR)}, 2015.

\bibitem[Kirby(2001)]{kirby2001spontaneous}
Simon Kirby.
\newblock Spontaneous evolution of linguistic structure-an iterated learning
  model of the emergence of regularity and irregularity.
\newblock \emph{IEEE Transactions on Evolutionary Computation}, 5\penalty0
  (2):\penalty0 102--110, 2001.

\bibitem[Kirby and Hurford(2002)]{kirby2002emergence}
Simon Kirby and James~R Hurford.
\newblock The emergence of linguistic structure: An overview of the iterated
  learning model.
\newblock \emph{Simulating the evolution of language}, pages 121--147, 2002.

\bibitem[Kirby et~al.(2015)Kirby, Tamariz, Cornish, and
  Smith]{kirby2015compression}
Simon Kirby, Monica Tamariz, Hannah Cornish, and Kenny Smith.
\newblock Compression and communication in the cultural evolution of linguistic
  structure.
\newblock \emph{Cognition}, 141:\penalty0 87--102, 2015.

\bibitem[Koehn(2004)]{koehn2004statistical}
Philipp Koehn.
\newblock Statistical significance tests for machine translation evaluation.
\newblock In \emph{Proc. of Empirical Methods in natural Language Processing
  (EMNLP)}, 2004.

\bibitem[Kokoska and Zwillinger(2000)]{kokoska2000crc}
Stephen Kokoska and Daniel Zwillinger.
\newblock \emph{CRC standard probability and statistics tables and formulae}.
\newblock Crc Press, 2000.

\bibitem[Kottur et~al.(2017)Kottur, Moura, Lee, and
  Batra]{kottur-etal-2017-natural}
Satwik Kottur, Jos{\'e} Moura, Stefan Lee, and Dhruv Batra.
\newblock Natural language does not emerge {`}naturally{'} in multi-agent
  dialog.
\newblock In \emph{Proc. of Empirical Methods in Natural Language Processing
  (EMNLP)}, 2017.

\bibitem[Krogh and Hertz(1991)]{krogh1991simple}
Anders Krogh and John Hertz.
\newblock A simple weight decay can improve generalization.
\newblock \emph{Advances in neural information processing systems}, 4, 1991.

\bibitem[Kuci{\'n}ski et~al.(2021)Kuci{\'n}ski, Korbak, Ko{\l}odziej, and
  Mi{\l}o{\'s}]{kucinski2021catalytic}
{\L}ukasz Kuci{\'n}ski, Tomasz Korbak, Pawe{\l} Ko{\l}odziej, and Piotr
  Mi{\l}o{\'s}.
\newblock Catalytic role of noise and necessity of inductive biases in the
  emergence of compositional communication.
\newblock \emph{Proc. of Advances in Neural Information Processing Systems
  (NeurIPS)}, 2021.

\bibitem[Lazaridou and Baroni(2020)]{lazaridou2020emergent}
Angeliki Lazaridou and Marco Baroni.
\newblock Emergent multi-agent communication in the deep learning era.
\newblock \emph{arXiv preprint arXiv:2006.02419}, 2020.

\bibitem[Lazaridou et~al.(2016)Lazaridou, Peysakhovich, and
  Baroni]{lazaridou2016multi}
Angeliki Lazaridou, Alexander Peysakhovich, and Marco Baroni.
\newblock Multi-agent cooperation and the emergence of (natural) language.
\newblock \emph{arXiv preprint arXiv:1612.07182}, 2016.

\bibitem[Lazaridou et~al.(2018)Lazaridou, Hermann, Tuyls, and
  Clark]{lazaridou2018emergence}
Angeliki Lazaridou, Karl~Moritz Hermann, Karl Tuyls, and Stephen Clark.
\newblock Emergence of linguistic communication from referential games with
  symbolic and pixel input.
\newblock In \emph{Proc. of International Conference on Learning
  Representations (ICLR)}, 2018.

\bibitem[Levenshtein et~al.(1966)]{levenshtein1966binary}
Vladimir~I Levenshtein et~al.
\newblock Binary codes capable of correcting deletions, insertions, and
  reversals.
\newblock In \emph{Soviet physics doklady}, volume 10:8, pages 707--710. Soviet
  Union, 1966.

\bibitem[Lewis(1969)]{lewis1969convention}
David~Kellogg Lewis.
\newblock \emph{Convention: A Philosophical Study}.
\newblock Cambridge, MA, USA: Wiley-Blackwell, 1969.

\bibitem[Li and Bowling(2019)]{li2019ease}
Fushan Li and Michael Bowling.
\newblock Ease-of-teaching and language structure from emergent communication.
\newblock In \emph{Proc. of Advances in Neural Information Processing Systems
  (NeurIPS)}, 2019.

\bibitem[Liu et~al.(2015)Liu, Luo, Wang, and Tang]{liu2015deep}
Ziwei Liu, Ping Luo, Xiaogang Wang, and Xiaoou Tang.
\newblock Deep learning face attributes in the wild.
\newblock In \emph{Proc. of the International Conference on Computer Vision
  (ICCV)}, 2015.

\bibitem[Lowe et~al.(2020)Lowe, Gupta, Foerster, Kiela, and Pineau]{Lowe2020On}
Ryan Lowe, Abhinav Gupta, Jakob Foerster, Douwe Kiela, and Joelle Pineau.
\newblock On the interaction between supervision and self-play in emergent
  communication.
\newblock In \emph{Proc. of International Conference on Learning
  Representations (ICLR)}, 2020.

\bibitem[Mnih et~al.(2016)Mnih, Badia, Mirza, Graves, Lillicrap, Harley,
  Silver, and Kavukcuoglu]{pmlr-v48-mniha16}
Volodymyr Mnih, Adria~Puigdomenech Badia, Mehdi Mirza, Alex Graves, Timothy
  Lillicrap, Tim Harley, David Silver, and Koray Kavukcuoglu.
\newblock Asynchronous methods for deep reinforcement learning.
\newblock In Maria~Florina Balcan and Kilian~Q. Weinberger, editors,
  \emph{Proceedings of The 33rd International Conference on Machine Learning},
  volume~48 of \emph{Proceedings of Machine Learning Research}, pages
  1928--1937, New York, New York, USA, 20--22 Jun 2016. PMLR.
\newblock URL \url{https://proceedings.mlr.press/v48/mniha16.html}.

\bibitem[Mu and Goodman(2021)]{mu2021emergent}
Jesse Mu and Noah Goodman.
\newblock Emergent communication of generalizations.
\newblock \emph{Proc. of Advances in Neural Information Processing Systems
  (NeurIPS)}, 2021.

\bibitem[Noukhovitch et~al.(2021)Noukhovitch, LaCroix, Lazaridou, and
  Courville]{noukhovitch2021emergent}
Michael Noukhovitch, Travis LaCroix, Angeliki Lazaridou, and Aaron Courville.
\newblock Emergent communication under competition.
\newblock In \emph{Proc. of International Conference on Autonomous Agents and
  MultiAgent Systems (AAMAS)}, 2021.

\bibitem[Oord et~al.(2018)Oord, Li, and Vinyals]{oord2018representation}
Aaron van~den Oord, Yazhe Li, and Oriol Vinyals.
\newblock Representation learning with contrastive predictive coding.
\newblock \emph{arXiv preprint arXiv:1807.03748}, 2018.

\bibitem[Ossenkopf et~al.(2022)Ossenkopf, Luck, and
  Mathewson]{ossenkopf2022which}
Marie Ossenkopf, Kevin~Sebastian Luck, and Kory~Wallace Mathewson.
\newblock Which language evolves between heterogeneous agents? - communicating
  movement instructions with widely different time scopes.
\newblock In \emph{Emergent Communication Workshop at ICLR 2022}, 2022.

\bibitem[Raviv et~al.(2019)Raviv, Meyer, and Lev-Ari]{raviv2019compo}
Limor Raviv, Antje Meyer, and Shiri Lev-Ari.
\newblock Compositional structure can emerge without generational transmission.
\newblock \emph{Cognition}, 182:\penalty0 151--164, 2019.
\newblock ISSN 0010-0277.
\newblock \doi{https://doi.org/10.1016/j.cognition.2018.09.010}.
\newblock URL
  \url{https://www.sciencedirect.com/science/article/pii/S0010027718302464}.

\bibitem[Ren et~al.(2020)Ren, Guo, Labeau, Cohen, and
  Kirby]{ren2020compositional}
Yi~Ren, Shangmin Guo, Matthieu Labeau, Shay~B. Cohen, and Simon Kirby.
\newblock Compositional languages emerge in a neural iterated learning model.
\newblock In \emph{Proc. of International Conference on Learning
  Representations (ICLR)}, 2020.

\bibitem[Resnick et~al.(2020)Resnick, Gupta, Foerster, Dai, and
  Cho]{resnick2019capacity}
Cinjon Resnick, Abhinav Gupta, Jakob Foerster, Andrew~M. Dai, and Kyunghyun
  Cho.
\newblock Capacity, bandwidth, and compositionality in emergent language
  learning.
\newblock In \emph{Proc. of Autonomous Agents and Multiagent Systems (AAMAS)},
  2020.

\bibitem[Rita et~al.(2020)Rita, Chaabouni, and Dupoux]{rita-etal-2020-lazimpa}
Mathieu Rita, Rahma Chaabouni, and Emmanuel Dupoux.
\newblock {``}{L}az{I}mpa{''}: Lazy and impatient neural agents learn to
  communicate efficiently.
\newblock In \emph{Proc. of the Conference on Computational Natural Language
  Learning (CoNLL)}, 2020.

\bibitem[Rita et~al.(2022)Rita, Strub, Grill, Pietquin, and Dupoux]{rita2022on}
Mathieu Rita, Florian Strub, Jean-Bastien Grill, Olivier Pietquin, and Emmanuel
  Dupoux.
\newblock On the role of population heterogeneity in emergent communication.
\newblock In \emph{Proc. of International Conference on Learning
  Representations (ICLR)}, 2022.

\bibitem[Russakovsky et~al.(2015)Russakovsky, Deng, Su, Krause, Satheesh, Ma,
  Huang, Karpathy, Khosla, Bernstein, et~al.]{russakovsky2015imagenet}
Olga Russakovsky, Jia Deng, Hao Su, Jonathan Krause, Sanjeev Satheesh, Sean Ma,
  Zhiheng Huang, Andrej Karpathy, Aditya Khosla, Michael Bernstein, et~al.
\newblock Imagenet large scale visual recognition challenge.
\newblock \emph{International journal of computer vision (IJCV)}, 115\penalty0
  (3):\penalty0 211--252, 2015.

\bibitem[Skyrms(2010)]{skyrms2010signals}
Brian Skyrms.
\newblock \emph{Signals: Evolution, learning, and information}.
\newblock OUP Oxford, 2010.

\bibitem[S{\l}owik et~al.(2002)S{\l}owik, Gupta, Hamilton, Jamnik, Holden, and
  Pal]{slowik2002exploring}
Agnieszka S{\l}owik, Abhinav Gupta, William~L Hamilton, Mateja Jamnik, Sean~B
  Holden, and Christopher Pal.
\newblock Exploring structural inductive biases in emergent communication.
\newblock \emph{arXiv preprint arXiv:2002.01335}, 2002.

\bibitem[Smith et~al.(2003)Smith, Kirby, and Brighton]{smith2003iterated}
Kenny Smith, Simon Kirby, and Henry Brighton.
\newblock Iterated learning: A framework for the emergence of language.
\newblock \emph{Artificial life}, 9\penalty0 (4):\penalty0 371--386, 2003.

\bibitem[Spike et~al.(2017)Spike, Stadler, Kirby, and Smith]{spike2017minimal}
Matthew Spike, Kevin Stadler, Simon Kirby, and Kenny Smith.
\newblock Minimal requirements for the emergence of learned signaling.
\newblock \emph{Cognitive science}, 41\penalty0 (3):\penalty0 623--658, 2017.

\bibitem[Srivastava et~al.(2014)Srivastava, Hinton, Krizhevsky, Sutskever, and
  Salakhutdinov]{JMLR:v15:srivastava14a}
Nitish Srivastava, Geoffrey Hinton, Alex Krizhevsky, Ilya Sutskever, and Ruslan
  Salakhutdinov.
\newblock Dropout: A simple way to prevent neural networks from overfitting.
\newblock \emph{Journal of Machine Learning Research}, 15\penalty0
  (56):\penalty0 1929--1958, 2014.
\newblock URL \url{http://jmlr.org/papers/v15/srivastava14a.html}.

\bibitem[Steels(1997)]{steels1997synthetic}
Luc Steels.
\newblock The synthetic modeling of language origins.
\newblock \emph{Evolution of communication}, 1\penalty0 (1):\penalty0 1--34,
  1997.

\bibitem[Sutton et~al.(2000)Sutton, McAllester, Singh, and
  Mansour]{sutton2000policy}
Richard~S Sutton, David~A McAllester, Satinder~P Singh, and Yishay Mansour.
\newblock Policy gradient methods for reinforcement learning with function
  approximation.
\newblock In \emph{Proc. of Advances in Neural Information Processing Systems
  (NIPS)}, 2000.

\bibitem[Szabó(2020)]{sep-compositionality}
Zoltán~Gendler Szabó.
\newblock {Compositionality}.
\newblock In Edward~N. Zalta, editor, \emph{The {Stanford} Encyclopedia of
  Philosophy}. Metaphysics Research Lab, Stanford University, {F}all 2020
  edition, 2020.

\bibitem[Tamariz and Kirby(2015)]{tamariz2015culture}
M{\'o}nica Tamariz and Simon Kirby.
\newblock Culture: copying, compression, and conventionality.
\newblock \emph{Cognitive science}, 39\penalty0 (1):\penalty0 171--183, 2015.

\bibitem[Townsend et~al.(2018)Townsend, Engesser, Stoll, Zuberb{\"u}hler, and
  Bickel]{townsend2018compositionality}
Simon~W Townsend, Sabrina Engesser, Sabine Stoll, Klaus Zuberb{\"u}hler, and
  Balthasar Bickel.
\newblock Compositionality in animals and humans.
\newblock \emph{PLoS Biology}, 16\penalty0 (8):\penalty0 e2006425, 2018.

\bibitem[Virtanen et~al.(2020)Virtanen, Gommers, Oliphant, Haberland, Reddy,
  Cournapeau, Burovski, Peterson, Weckesser, Bright, {van der Walt}, Brett,
  Wilson, Millman, Mayorov, Nelson, Jones, Kern, Larson, Carey, Polat, Feng,
  Moore, {VanderPlas}, Laxalde, Perktold, Cimrman, Henriksen, Quintero, Harris,
  Archibald, Ribeiro, Pedregosa, {van Mulbregt}, and {SciPy 1.0
  Contributors}]{2020SciPy-NMeth}
Pauli Virtanen, Ralf Gommers, Travis~E. Oliphant, Matt Haberland, Tyler Reddy,
  David Cournapeau, Evgeni Burovski, Pearu Peterson, Warren Weckesser, Jonathan
  Bright, St{\'e}fan~J. {van der Walt}, Matthew Brett, Joshua Wilson, K.~Jarrod
  Millman, Nikolay Mayorov, Andrew R.~J. Nelson, Eric Jones, Robert Kern, Eric
  Larson, C~J Carey, {\.I}lhan Polat, Yu~Feng, Eric~W. Moore, Jake
  {VanderPlas}, Denis Laxalde, Josef Perktold, Robert Cimrman, Ian Henriksen,
  E.~A. Quintero, Charles~R. Harris, Anne~M. Archibald, Ant{\^o}nio~H. Ribeiro,
  Fabian Pedregosa, Paul {van Mulbregt}, and {SciPy 1.0 Contributors}.
\newblock {{SciPy} 1.0: Fundamental Algorithms for Scientific Computing in
  Python}.
\newblock \emph{Nature Methods}, 17:\penalty0 261--272, 2020.
\newblock \doi{10.1038/s41592-019-0686-2}.

\bibitem[Wagner et~al.(2003)Wagner, Reggia, Uriagereka, and
  Wilkinson]{wagner2003progress}
Kyle Wagner, James~A Reggia, Juan Uriagereka, and Gerald~S Wilkinson.
\newblock Progress in the simulation of emergent communication and language.
\newblock \emph{Adaptive Behavior}, 11\penalty0 (1):\penalty0 37--69, 2003.

\bibitem[Williams and Peng(1991)]{williams1991function}
Ronald~J Williams and Jing Peng.
\newblock Function optimization using connectionist reinforcement learning
  algorithms.
\newblock \emph{Connection Science}, 3\penalty0 (3):\penalty0 241--268, 1991.

\bibitem[Wray and Grace(2007)]{wray2007543}
Alison Wray and George~W. Grace.
\newblock The consequences of talking to strangers: Evolutionary corollaries of
  socio-cultural influences on linguistic form.
\newblock \emph{Lingua}, 117\penalty0 (3):\penalty0 543--578, 2007.
\newblock ISSN 0024-3841.
\newblock The Evolution of Language.

\bibitem[Zaslavsky et~al.(2018)Zaslavsky, Kemp, Regier, and
  Tishby]{zaslavsky2018}
Noga Zaslavsky, Charles Kemp, Terry Regier, and Naftali Tishby.
\newblock Efficient compression in color naming and its evolution.
\newblock \emph{Proc. of the National Academy of Sciences}, 115\penalty0
  (31):\penalty0 7937--7942, 2018.
\newblock \doi{10.1073/pnas.1800521115}.
\newblock URL \url{https://www.pnas.org/doi/abs/10.1073/pnas.1800521115}.

\bibitem[Łukasz Kuciński et~al.(2020)Łukasz Kuciński, Kołodziej, and
  Milos]{Kuciski2020EmergenceOC}
Łukasz Kuciński, P.~Kołodziej, and Piotr Milos.
\newblock Emergence of compositional language in communication through noisy
  channel.
\newblock In \emph{Language in Reinforcement Learning (LaReL), ICML Workshop},
  2020.

\end{thebibliography}
